\documentclass[english,aps,prb, twocolumn, superscriptaddress,showpacs, amsmath]{revtex4-2}
\usepackage[T1]{fontenc}
\usepackage{xcolor}
\usepackage{units}
\usepackage{amssymb}
\usepackage{graphicx}
\usepackage{esint}
\usepackage{bm}
\usepackage{natbib}
\usepackage{amsmath}
\usepackage[colorlinks=true, citecolor={blue!80!black}, urlcolor={blue!50!black}, linkcolor = {blue!80!black}]{hyperref}
\usepackage{braket}
\usepackage{mathtools}
\usepackage{physics}
\usepackage{soul}

\setcitestyle{journalcolor= blue}

\usepackage{babel}

\begin{document}

\title{
Ground state phase diagram and ``parity flipping'' microwave transitions \\in a gate-tunable Josephson Junction}
\newcommand{\affA}{\affiliation{Departamento de F\'{\i}sica Te\'orica de la Materia Condensada, \mbox{Condensed Matter Physics Center (IFIMAC)} and Instituto Nicol\'as Cabrera, Universidad Aut\'onoma de Madrid, 28049 Madrid, Spain}}
\newcommand{\affB}{\affiliation{Quantronics group, Service de Physique de l'\'Etat Condens\'e \mbox{(CNRS, UMR 3680)}, IRAMIS, CEA-Saclay, Universit\'e Paris-Saclay, 91191 Gif-sur-Yvette, France}}
\newcommand{\affC}{\affiliation{Centro At\'omico Bariloche and Instituto Balseiro, CNEA, CONICET, 8400 San Carlos de Bariloche, R\'io Negro, Argentina}}
\newcommand{\affF}{\affiliation{NNF Quantum Computing Programme, Niels Bohr Institute, University of Copenhagen, Universitetsparken 5, 2100 Copenhagen, Denmark}}
\newcommand{\affE}{\affiliation{Center for Quantum Devices, Niels Bohr Institute, University of Copenhagen, Universitetsparken 5, 2100 Copenhagen, Denmark}}
\newcommand{\eqContrib}{\thanks{These authors contributed equally to this work.}}

\author{M. R. Sahu}
\eqContrib
\affB
\author{F. J. \surname{Matute-Ca\~nadas}}
\eqContrib
\affA
\author{M. Benito}
\affB
\author{P. Krogstrup}
\affF\affE
\author{J.~Nyg\r{a}rd}
\affE
\author{M. F. Goffman}
\affB
\author{C. Urbina}
\affB
\author{A. \surname{Levy Yeyati}}
\affA
\author{H. Pothier}
\email[Corresponding author: ]{hugues.pothier@cea.fr}
\affB

\date{\today}

\begin{abstract}
  We probed a gate-tunable InAs nanowire Josephson weak link by coupling it to  a microwave resonator. Tracking the resonator frequency shift when the weak link is close to pinch-off, we observe that the ground state of the latter alternates between a singlet and a doublet when varying either the gate voltage or the superconducting phase difference across it.
  The corresponding microwave absorption spectra display lines that approach zero energy close to the singlet-doublet boundaries, suggesting parity flipping transitions, which are in principle forbidden in microwave spectroscopy and expected to arise only in tunnel spectroscopy.
  We tentatively interpret them by means of an ancillary state isolated in the junction acting as a reservoir for individual electrons.
\end{abstract}

\maketitle

\section{Introduction}

 Particle number parity effects are widespread in mesoscopic superconductivity \cite{Averin_Nazarov_1993,nazarov_blanter_2009}. They first appeared in circuits containing small metallic superconducting islands \cite{Tuominen1992,Lafarge_free_energy_1993,Lafarge_2eQuantization1993,Joyez_1994,Ralph1995} and in semiconductor-superconductor hybrids, like a quantum dot coupled to superconducting electrodes through tunnel barriers \cite{DeFranceschi2010}. In these systems, the electrodynamics depends crucially on charging effects in the island or the quantum dot \cite{IngoldNazarov1992}. More recently, parity effects were shown to arise in mesoscopic Josephson weak links, structures containing no island or quantum dots and therefore no significant charging energy. Here, the physics is understood in terms of a few Andreev bound states (ABS), subgap localized quasiparticle states with energies governed by the superconducting phase difference across the weak link. The odd or even many-body occupations of these states result in markedly different weak link electrodynamic properties. They are probed using microwave circuit-QED (cQED) techniques, microwave absorption spectroscopy, quasiparticle addition spectroscopy, and combinations of them \cite{janvier2015coherent,Bretheau2013,pillet2010,Wesdorp2021}. In the case of infinitely short weak links, realized with atomic contacts between two superconducting leads \cite{vanRuitenbeek1996}, all the observed features are explained in terms of non-interacting junction models \cite{Tagliacozzo2019}. There is also a wealth of experimental results on gate-tunable, finite-length weak links, based on semiconducting nanowires and which are also described in terms of ABS \cite{Lee2013, vanWoerkom2017,Hays2018,Tosi2019,Hays2021,Prada2020}. 
There is recent evidence that in these weak links, even with well transmitted conduction channels, interactions do play a role, albeit just as a small perturbation \cite{matute2022signatures,fatemi2022microwave}. When approaching pinch-off in the same devices, one expects quantum dot physics to become relevant and influence the parity dynamics.
 
Here we present circuit-QED measurements on gate-tunable InAs nanowire weak links  \cite{Krogstrup2015} close to pinch-off. We observe, both as a function of gate voltage and phase difference, features that we associate to transitions between ground states of different parity, similarly to what is observed in quantum dots. Remarkably, the corresponding microwave absorption spectra exhibit transition lines that as a function of gate voltage bear a close resemblance with those typically observed in an addition spectrum \cite{pillet2010,goffman2013}, and therefore seem to couple states of different parity, a forbidden process in photon absorption spectroscopy. We interpret these results as revealing the presence of an ancillary, weakly coupled quantum level, which allows mimicking parity transitions on the main transport channel without a change in the global parity.

\section{Basic concepts}

\begin{figure}[t]
\begin{center}
\includegraphics[width=0.47\textwidth]{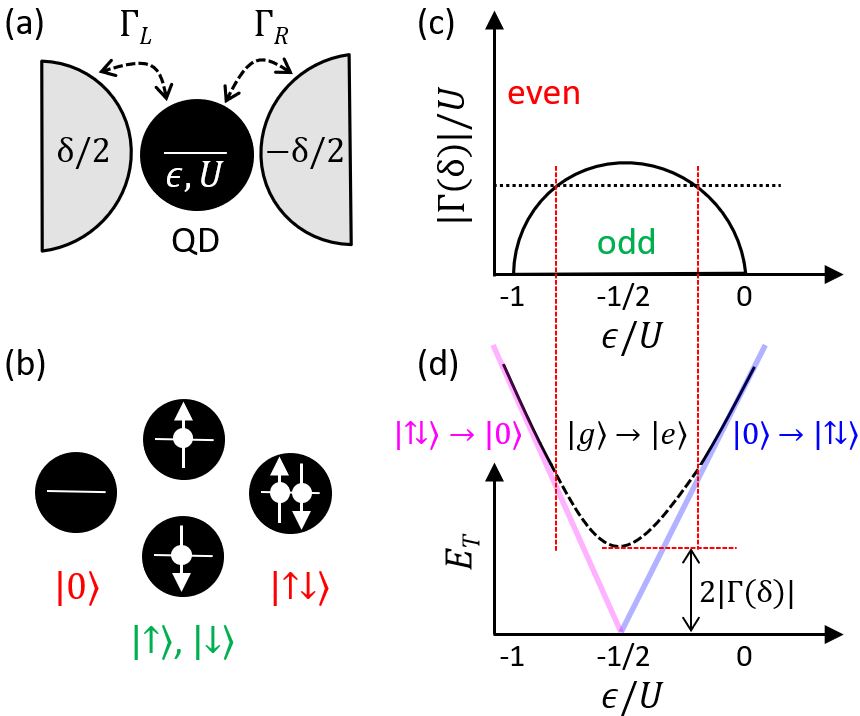}
 \caption{(\textbf{a}) Scheme of the superconducting single level Anderson  model.  (\textbf{b}) The four states correspond to the possible occupancies of the quantum dot level. (\textbf{c}) Phase diagram for model in the infinite gap limit. (\textbf{d}) Black curve represents the energy $E_T$ of the transition $\ket{g} \rightarrow \ket{e}$, with $\ket{g}$ and $\ket{e}$ the ground and excited states in the even parity sector, for a given $\Gamma(\delta)$ represented as a horizontal dotted line in (c). The dashed part of the curve should not be visible in the zero temperature limit as it corresponds to the odd ground state region. States \(\ket{g}\) and \(\ket{e}\) are linear combinations of the dot states \(\ket{0}\) and \(\ket{\uparrow\downarrow}\), hybridized by the effective pairing \(\Gamma(\delta)\). Blue and magenta straight lines correspond to the limiting cases when \(\Gamma(\delta) = 0\).}
\label{fig1}
\end{center}
\end{figure}

 When a few-channel conductor connects two superconductors in a phase-biased configuration, various regimes are encountered depending on the relative size of the coupling to the leads $\Gamma$, the Coulomb repulsive energy $U$, and the superconducting gap $\Delta$ \cite{nazarov_blanter_2009,martin2011josephson}. In the limit of large coupling, the system is well described by electrons and holes bouncing back and forth between the electrodes, with Andreev reflections at each interface, giving rise to supercurrent-carrying Andreev bound states \cite{andreev1966electron,kulik1969macroscopic,furusaki1991current}. The opposite limit is that of a quantum dot weakly coupled to the superconducting leads, usually described using a single-level Anderson impurity model \cite{anderson1963probable,glazman1989resonant,choi2004kondo,oguri2004quantum,tanaka2007kondo,martin2011josephson,karrasch2008josephson,kadlecova2019practical,meden2019anderson,Kurilovich2021,Hermansen2022}, as schematically presented in Fig.~\ref{fig1}(a). 
The energy $\epsilon$ of the dot level (referred to the leads' Fermi level) can be tuned by means of an electrostatic gate. In the absence of a magnetic field, the level is spin-degenerate, and the four possible dot states $\ket{0}$, $\ket{\uparrow},$ $\ket{\downarrow}$ and $\ket{\uparrow\downarrow}$ shown in Fig.~\ref{fig1}(b), characterized by the level occupation, split into two categories. When 
the dot occupancy is even (states $\ket{0}$ or $\ket{\uparrow\downarrow}$), the corresponding energies are 0 and $2\epsilon+U$. When the occupancy is odd, the spin-degenerate state $\ket{\sigma}$ has energy $\epsilon$. One finds that the ground state of the dot is odd if $-1 < \epsilon /U < 0$, and it is even  otherwise (see Fig.~\ref{fig1}(c)). Increasing the coupling to the superconducting leads favors a singlet superposition of the even states, $\ket{g}$, while gradually diminishing the extension of the odd (doublet) ground state in the phase diagram. This effect is most simply captured in the infinite gap limit, where the effective pairing is given by $\Gamma(\delta)=\Gamma_L e^{i\delta/2}+\Gamma_Re^{-i\delta/2}$, with $\Gamma_{L,R}$ the tunneling rates to the left and right leads and $\delta$ the superconducting phase difference between the two leads \cite{martin2011josephson,Meng2009}. This behavior has been investigated in a number of works \cite{pillet2010,deacon2010tunneling,goffman2013,chang2013tunneling,lee2014spin,li20170,lee2017scaling}, and revived by recent experiments using microwave techniques \cite{fatemi2022microwave, bargerbos2022singlet,Bargerbos2022doublet}. Finally, panel (d) shows the $\epsilon$ dependence of the transition energy from the even ground state $\ket{g}$ towards the even excited state $\ket{e}$. 

\begin{figure*}[t]
\begin{center}
\includegraphics[width=1\textwidth]{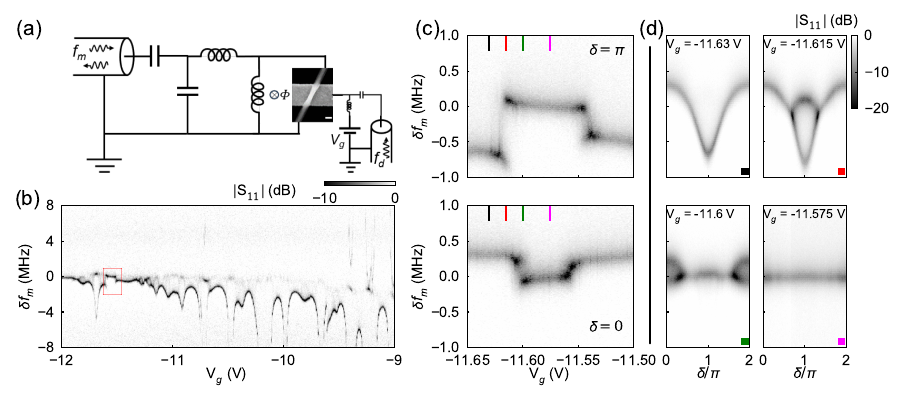}
 \caption{(\textbf{a}) Schematic of the measurement setup with SEM image (scale bar, 200 nm) of the InAs nanowire weak link. $f_m$ is the measurement frequency, $V_g$ the gate voltage and $f_d$ the drive frequency applied to the gate electrode through a bias tee in two-tone spectroscopy measurements. $\Phi$ is the magnetic flux related to phase difference $\delta$ by $\delta = 2\pi \Phi / \Phi_0,$ with $\Phi_0=h/2e$ the flux quantum. (\textbf{b}) 2D grey scale map of the amplitude of reflection coefficient $|S_{11}|$ plotted as a function of $\delta f_m$ $=f_m-f_0$ and $V_g$, at phase difference $\delta$ = $\pi$. (\textbf{c}) Upper panel: A higher resolution 2D grey scale map of the single-tone spectrum $|S_{11}|(V_g,\delta f_m)$ at $\delta$ = $\pi$ in the highlighted region of (b); Lower panel: corresponding 2D map of $|S_{11}|(V_g,\delta f_m)$ at $\delta$ = $0$. (\textbf{d}) 2D grey scale map of $|S_{11}|$ plotted as a function of gate voltage and phase difference ($\delta / \pi$) at several gate voltages ($V_g = -11.63$,$-11.615$,$-11.6$ and $-11.575$~V) marked by vertical ticks in (c) sharing the same color as the squares at the bottom right of the corresponding panels in (d).
}
\label{fig2}
\end{center}
\end{figure*} 

\section{Experimental results}

The experimental setup is schematized in Fig.~\ref{fig2}(a). An InAs nanowire weak link is placed in a superconducting loop threaded by a magnetic flux $\Phi$. The phase difference $\delta$ across the weak link is given by $\delta=2\pi\Phi/\Phi_0,$ with $\Phi_0=h/2e$ the flux quantum. The wire is suspended over a metallic gate, biased at voltage $V_g$, which allows to control the electron density. The loop participates in the inductance of a quarter-wavelength coplanar wave-guide resonator made out of NbTiN. The occupation of the Andreev states in the nanowire is inferred from the (complex) reflection coefficient of a microwave tone at frequency $f_m$ close to the bare resonance frequency $f_0=7.00~$GHz of the resonator measured when the weak link is fully depleted. The total quality factor of the resonator is $Q=23000$.
A second tone (``drive'' tone) with frequency $f_d$, applied through the gate line, allows probing the absorption excitation spectrum of the weak link. A detailed discussion of measurement setup and device fabrication is presented in the supplemental material (SM) \cite{SM}.

First, we present the single-tone measurements of the reflection coefficient $S_{11}$. 
The amplitude $|S_{11}(\delta f_m)|,$ where $\delta f_m = f_m-f_0,$ is presented as a function of $V_g$ in Fig.~\ref{fig2}(b). Dark lines mark minima of $|S_{11}|$ associated with the resonance frequency of the resonator modified by the occupation of Andreev states in the weak link. Highly dispersing lines can be related to pair transitions with gate-modulated transition energies \cite{Metzger2021,
Park2020,Kurilovich2021,Hermansen2022}.
When $V_g$ approaches $-12~$V, the oscillations of the resonance frequency fade away, marking the complete depletion of the nanowire. All along the scan, one also observes a weak resonance at $\delta f_m \approx 0,$ which corresponds to a state very weakly coupled to the resonator \cite{Metzger2021}.

Unique jumps in the resonance frequency are observed close to pinch-off in the single-tone data in Fig.~\ref{fig2}(b). One such region is highlighted by a red rectangle around $V_g=-11.6$~V. A higher resolution measurement of single-tone spectra around this highlighted region at $\delta=\pi$ is shown in the upper panel of Fig.~\ref{fig2}(c). Similar jumps are observed for $\delta=0$ as shown in the lower panel of Fig.~\ref{fig2}(c). The central plateaus in both plots correspond to resonance frequencies very close to the bare resonance frequency $f_0$, whereas the outer regions appear at $f_m < f_0$ for $\delta = \pi$ and at $f_m > f_0$ for $\delta=0$. To better understand these behaviors, 2D grey scale maps of $|S_{11}|$ as a function of phase difference $\delta$ and frequency $\delta f_m$ are plotted in Fig.~\ref{fig2}(d), at several gate voltages. At $V_g = -11.63$~V (top left), we observe a single transition frequency strongly dispersing with phase, which is expected for the supercurrent-carrying even (singlet) ground state in the single-level Anderson model. In contrast, at $V_g = -11.575$~V (bottom right), the resonance frequency is almost phase-independent and lies very close to $\delta f_m = 0$, close to behavior of an odd (doublet) ground state with suppressed supercurrent. From these observations, we infer that a strong signal on the central plateau of the resonances in Fig.~\ref{fig2}(c) corresponds to an odd-like ground state, and that a strong signal on the outer regions corresponds to an even-like ground state. 
As can be seen from the top-right and bottom-left panels of Fig.~\ref{fig2}(d), at intermediate gate voltages, we observe either one or two resonance frequencies depending on phase, which indicate a phase diagram of the singlet/doublet ground states that not only depends on $V_g$, but also on the phase difference $\delta$ \cite{bargerbos2022singlet,fatemi2022microwave}.

\begin{figure}[t!]
\begin{center}
\includegraphics[width=0.5\textwidth]{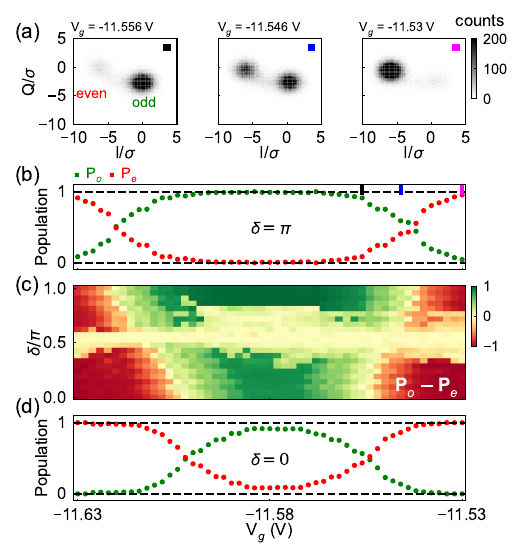}
 \caption{(\textbf{a}) Histogram of 50000 measurements of $S_{11}$ at a fixed measurement frequency close to $f_0$ in the IQ plane, at gate voltages $V_g=-11.556,$ $-11.546$, and $-11.53$~V, respectively. Each measurement produce mean-I and mean-Q over a 500 ns measurement duration. (\textbf{b}) Population of odd-like state, $P_o$ (green)  and even-like state, $P_e$ (red) are plotted as a function of $V_g$ at $\delta = \pi$. (\textbf{c}) 2D color map of polarization, $P_o - P_e$, plotted as a function of $V_g$ and $\delta / \pi$. (\textbf{d}) $P_o$ and $P_e$ plotted as a function of $V_g$ at $\delta = 0$.
}
\label{fig3}
\end{center}
\end{figure} 

\begin{figure*}[t]
\begin{center}
\includegraphics[width=1\textwidth]{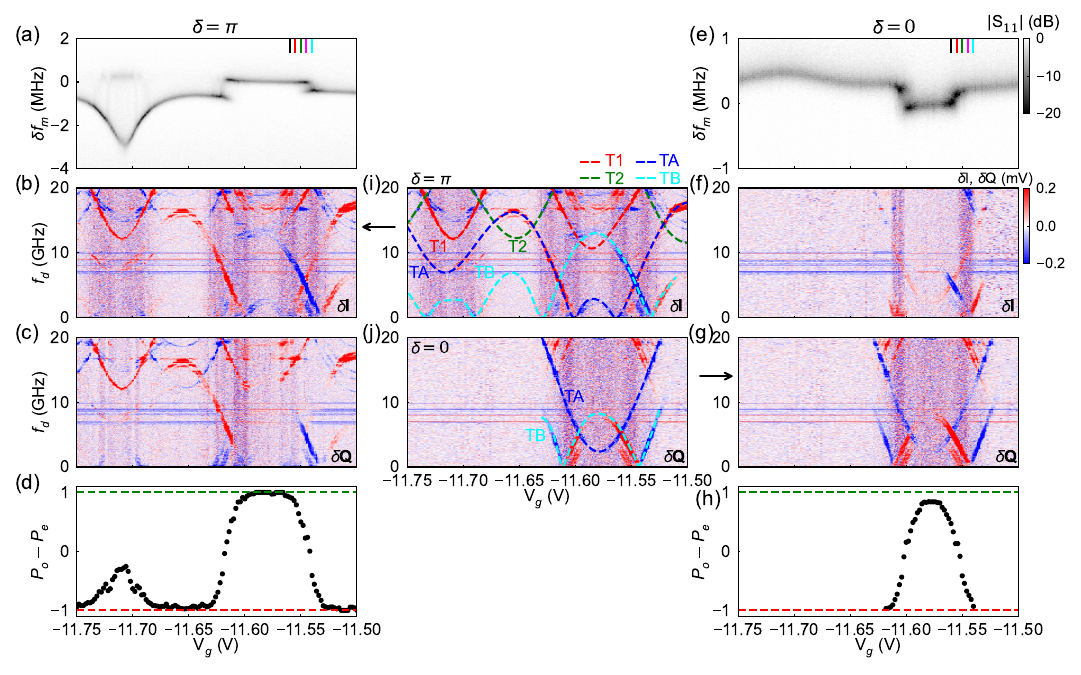}
 \caption{(\textbf{a}) 2D grey scale map of single-tone spectrum $|S_{11}|(V_g,\delta f_m)$ at $\delta = \pi$. (\textbf{b-c}) 2D color map of $\delta I$ and $\delta Q$ component of the two-tone spectroscopy, respectively, plotted as a function of drive frequency $f_d$ and $V_g$, at $\delta = \pi$. (\textbf{d}) Polarization, $P_o - P_e$, plotted as a function of $V_g$. Polarization=-1 (red dashed line) imply fully even-like ground state, polarization=1 (green dashed line) implies fully odd-like ground state. (\textbf{e-h}) Same as (a-d) at $\delta = 0$. (\textbf{i}) Duplicate of (b) where the four transitions T1 (red), T2 (green), TA (blue), and, TB (cyan) are highlighted by dashed lines at $\delta = \pi$. (\textbf{j}) Duplicate of (g) where the TA (blue) and TB (cyan) are highlighted by dashed lines at $\delta = 0$.
}
\label{fig4}
\end{center}
\end{figure*}

The single-tone results shown in Fig.~\ref{fig2} are measurements of $S_{11}$ averaged over a long (33~$\mu$s) duration at a given $f_m$. It reflects the different values, corresponding to different ABS occupations, taken by $S_{11}$ during the averaging time. 
Information about the ABS occupation dynamics can be accessed by performing a series of successive short measurements of $S_{11}$. 
We performed 50000 measurements of $S_{11}$ at a frequency close to $f_0$, with a time per point of 500~ns, each measurement producing a mean value of real (in-phase, $I$) and imaginary (quadrature, $Q$) components. Using these 50000 $I$ and $Q$ data, we plot histograms in the $IQ$ plane as shown in Fig.~\ref{fig3}(a) at three settings of $V_g$ around one of the region where we observe the jump in resonance frequency in Fig.~\ref{fig2}c. We observe two clouds in the $IQ$ plane, which correspond to the lower energy even and odd states of the weak link. By using a Gaussian mixture model (GMM) \cite{scikit-learn} we extracted the population of the two states ($P_o$ and $P_e$ correspond to populations of the odd- and even-like states, respectively) as a function of $V_g$ at $\delta$ = $\pi$ (Fig.~\ref{fig3}(b)), and at $\delta=0$ (Fig.~\ref{fig3}(d)). 
The $V_g$ region for which the odd-like  state is observed is larger at $\delta$ = $\pi$ compared to $\delta = 0$. 
In Fig.~\ref{fig3}(c), we show the 2D color map of polarization, $P_o-P_e$, as a function of $V_g$ and $\delta$, showing the full phase diagram of the singlet--doublet phase transition. In the region around $\delta / \pi \sim 0.5$ the clouds overlap and GMM prediction does not work. The procedure also fails when only one state is visible (strong polarization).

Lifetimes of the singlet and doublet states can be evaluated by performing a continuous version of the above measurement, which is presented in detail in the Supplemental Material. When the population of one of the states is close to 1, we observed its lifetime to be order of a milli-second with the lifetime of the other state being few micro seconds, similar to earlier experiment \cite{bargerbos2022singlet}.

The observations from Fig.~\ref{fig2} and Fig.~\ref{fig3} can be qualitatively understood by the fact that close to the pinch-off the coupling of the weak link to the superconducting leads can be significantly reduced, so that it behaves like a quantum dot. The system can then be modeled by a single level Anderson model, which in the infinite gap limit produces the phase diagram of singlet--doublet ground states shown in Fig.~\ref{fig1}(c) \cite{Meng2009,Hermansen2022}. In our experiment, the gate voltage mainly tunes the position of the energy level, whereas the phase difference between the superconducting leads tunes the effective coupling \(\Gamma(\delta)  =\Gamma_L e^{i\delta/2} + \Gamma_R e^{-i\delta/2}\). The fact that $\Gamma_\pi =\Gamma_L - \Gamma_R$ is lower in magnitude than $\Gamma_0 =\Gamma_L + \Gamma_R$ explains why the doublet is observed over a larger range of $V_g$ at $\delta = \pi$ that at $\delta = 0$.

We now present the two-tone spectroscopy results, which are measurements of the change in the reflection coefficient $S_{11}$ at a fixed frequency $f_m$ in presence of a drive tone with variable frequency $f_d$. For a given $f_d$, the $I$ and $Q$ components of $S_{11}$ are measured both when the drive is on and off, and the differences $\delta I$ and $\delta Q$ are recorded. 
In Fig.~\ref{fig4}(b-c) we show the 2D color map of $\delta I$ and  $\delta Q$ as a function of $V_g$ and $f_d$, at phase difference $\delta = \pi$. We use the the 2D color map of $\delta I$ in Fig.~\ref{fig4}(i) to highlight with dashed curves the four transitions that we will be discussing in the following. Transitions T1 and T2 (red and green) have rounded minima at finite frequency. This behavior is generic in Andreev nanowire weak links \cite{Tosi2019,thesisMetzger}.
In contrast, transitions TA and TB, which reach zero frequency (within experimental accuracy) with cusps, are anomalous and we observe them only near pinch-off. They resemble tunneling spectroscopy data in superconducting quantum dots, where they are associated to a quantum phase transition between even and odd ground state \cite{pillet2010}. 
In addition, TA and TB correspond to population transfer between the even and odd clouds shown in Fig.~\ref{fig3}(a) (more details in SM \cite{SM}). As will be discussed in Section III, we could reproduce them by introducing an ancillary level to the single-level Anderson model. In Fig.~\ref{fig4}(a) and Fig.~\ref{fig4}(d) we plot the single-tone spectrum and polarization $P_o-P_e$, respectively, as a function of $V_g$ at $\delta = \pi$. Interestingly, TA and TB intersect at the gate voltages very close to the singlet to doublet phase transition points \textit{i.e.} around the gate voltages where the polarization changes sign. 

When changing the phase from $\pi$ to 0, line TB changes but still exhibits cusps at zero frequency, while TA shifts up and does not reach zero frequency any longer, as shown in Fig.~\ref{fig4}(e-h) and Fig.~\ref{fig4}(j). 

\begin{figure}[t!]
\begin{center}
\includegraphics[width=0.5\textwidth]{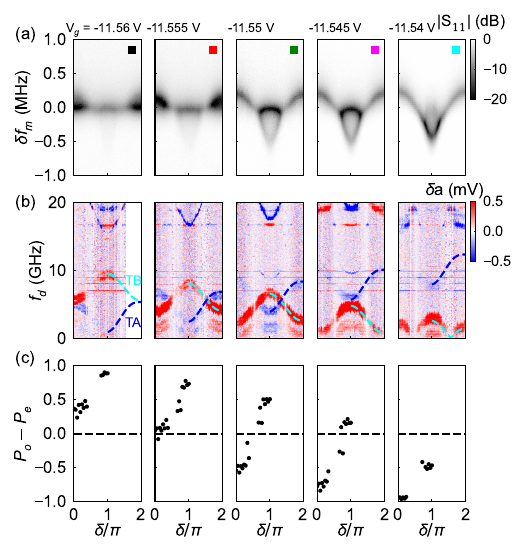}
 \caption{(\textbf{a}) 2D color maps of single-tone spectrum $|S_{11}|(\delta,\delta f_m)$ at gate voltages $V_g = -11.56$, $-11.555$, $-11.55$, $-11.545$ and $-11.54$~V, respectively (marked as colored vertical lines in Fig.~3(b) and Fig.~3(c)). (\textbf{b}) 2D color maps of two-tone spectrum $\delta a(\delta,f_d)$. (\textbf{c}) Polarization $P_o-P_e$ as a function of $\delta$, measured between 0 and $\pi$. Missing data at certain values of $\delta$ correspond to situations where GMM prediction does not work.}
 \label{fig5}
 \end{center}
\end{figure} 
 
Now, we present the phase dependence of the two-tone spectra at several gate voltages in Fig.~\ref{fig5}(b), together with the corresponding single-tone measurements in Fig.~\ref{fig5}(a) and polarization in Fig.~\ref{fig5}(c). The color code of the 2D color maps in Fig.~\ref{fig5}(b) represents the amplitude ($\delta a$) of the shift of $S_{11}$ in a $\delta$ dependent phase direction in the $IQ$ plane, such that the contrast of TA is maximized (more details in SM \cite{SM}). In the right half of each panel of Fig.~\ref{fig5}(b), the TA and TB are highlighted with dashed lines with same colors as Figs.~\ref{fig4}(b) and \ref{fig4}(f), \textit{i.e.} blue and cyan, respectively.
For the five gate voltages shown in Fig.~\ref{fig5}, the single-tone data in Fig.~\ref{fig5}(a) as well as the corresponding population data in Fig.~\ref{fig5}(c) show the gradual shrinking of the region for which the odd-like ground state is observed. The spacing between the crossing points of TA and TB also follows a similar decreasing trend. These observations are consistent with the theoretical model discussed below.

\section{Ancillary level model}

We now focus on the two lowest transition lines (TA and TB) versus gate voltage in the range where they exhibit cusps with cusps close to zero drive frequency (Fig.~\ref{fig4}). A possible explanation for these cusps is the occurrence of replicas involving the absorption or emission of a resonator photon with energy $hf_r$, thus appearing at energies \(E_T\pm hf_r\), where \(E_T\) is the bare transition energy. In the case where $E_T<hf_r$ and there is a significant population in the excited weak link state, there would also appear transition lines with energy \(-E_T+hf_r\) corresponding to the excitation of a resonator photon with relaxation in the weak link. This set of replica lines gives rise to cusps when $E_T$ crosses $hf_r.$
However, in our case, the replica mechanism should be discarded for the following reasons. On the one hand, if one of these anomalous transitions were a replica of the other one (as their constant vertical separation \({\sim}11\)~GHz in frequency suggests), one would still need to explain the appearance of cusps in the {\it other}  anomalous line. Moreover, the constant shift \({\sim} 11\)~GHz should correspond to one spurious resonator mode, which is not visible in the spectra. On the other hand, the anomalous lines cannot be replicas of transition lines T1 and T2, as illustrated in Fig.~S9 of the SM \cite{SM}. Ultimately, the highly symmetric disposition of the cusps and their proximity with the singlet/doublet boundaries hint at a different mechanism.

\begin{figure}[h]
\begin{center}
\includegraphics[width=0.475\textwidth]{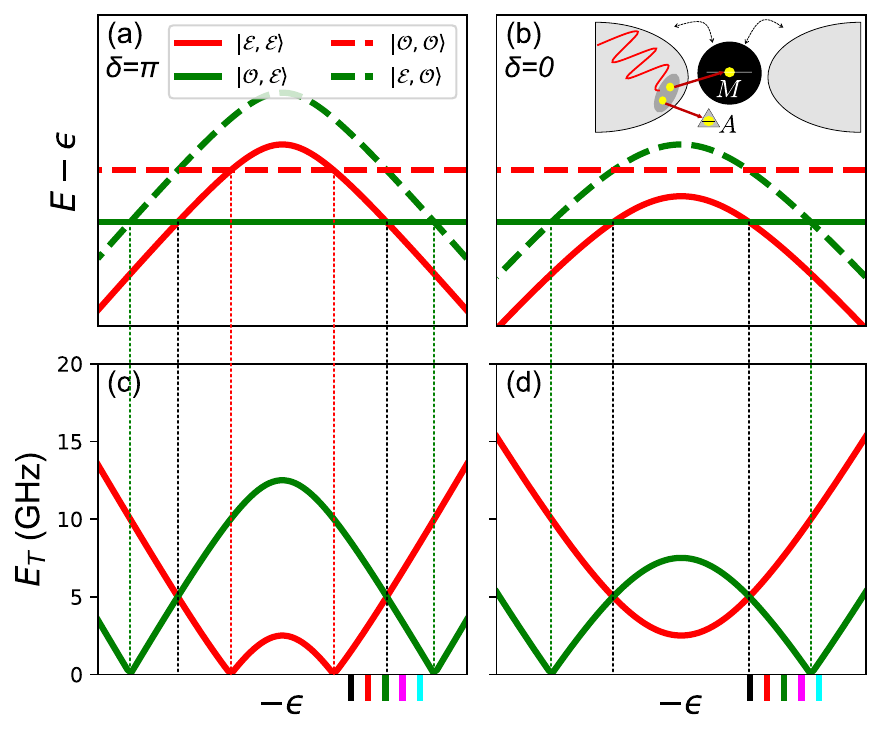}
 \caption{Upper row: Lowest many-body energies of the ancillary level model over the position \(\epsilon\) of the main channel level, at phase differences \(\delta{=}\pi\) (a) and \(0\) (b). Energies are plotted with a global shift of \(\epsilon\), and indicate even (odd) global parity with red (green) colour, and empty (filled) occupation in the ancillary level with solid (dashed) lines (second entry of ket \(\ket{M,A}\)). Inset in (b) represents the mechanism of local parity flip. Lower row:  corresponding global parity conserving transitions. Vertical dotted lines are placed at the singlet/doublet boundaries in the main channel (black) and at the crossings between states with the same global parity (red/green). Parameters are chosen to qualitatively reproduce the two-tone measurements in Fig.~\ref{fig4} around the gate range \(V_g\in [-11.65, -11.50] \textrm{V}\) (see SM): $\Gamma_L=2.5$,$\Gamma_R=8$,$\epsilon_A=5$,$U=26$ (units in GHz).
}
 \label{fig:theo_gate}
 \end{center}
\end{figure}

Indeed, when just one line is considered, its cusps close to the singlet/doublet boundaries suggest that it connects the singlet with the doublet states, as their energy difference crosses zero at the boundaries. Connecting singlet to doublet states when exciting with microwaves is forbidden because parity should be conserved. However, the situation changes if an ancillary dot level \(A\), weakly coupled to the main channel \(M\), is  added to the model (see inset in Fig.~\ref{fig:theo_gate}b). This configuration allows to explain the other line as well and has already been used to describe some transport experiments in semiconducting nanowire Josephson junctions \cite{Levajac2023}. We model it with a Hamiltonian $H = H_M + H_A$ such that

\begin{equation}
	H_M = \sum_{\sigma} \epsilon d^\dagger_\sigma d_\sigma  
	+ U n_\uparrow n_\downarrow
	+ \left( \Gamma(\delta) d^\dagger_\uparrow d^\dagger_\downarrow + h.c. \right),
\end{equation}
where $d^{\dagger}_{\sigma}$ creates an electron with spin $\sigma$ on the dot and $n_{\sigma} = d^{\dagger}_{\sigma}d_{\sigma}$, corresponds to the infinite gap Anderson model described above; and
\begin{equation}
	H_A = \sum_{\sigma} \epsilon_A d^\dagger_{A\sigma} d_{A\sigma}  
	+ U_A n_{A\uparrow} n_{A\downarrow}
	+ \left(t_A d^\dagger_{A\sigma} d_{\sigma} + h.c. \right)
\end{equation}
describes the ancillary dot level $\epsilon_A$, weakly coupled to the main channel by a vanishing tunnel amplitude $t_A$ and endowed with a charging energy \(U_A\) that forbids its double occupation.

When restricted to the main channel, the lowest energy levels in each parity sector even (\(\mathcal{E}\)) and odd (\(\mathcal{O}\)) are

\begin{align}
	 \ket{\mathcal{E}}_M &= u\ket{0}_M - v e^{i\theta}\ket{\uparrow\downarrow}_M  &\rightarrow\quad&   E_{\mathcal{E}} = \xi - \sqrt{\xi^2+|\Gamma(\delta)|^2} \nonumber \\
	 \ket{\mathcal{O}}_M &= \ket{\sigma}_M   &\rightarrow\quad& E_{\mathcal{O}} = \epsilon,
\end{align}

\noindent with \(\xi = \epsilon + U/2\), \(u(v) = \frac{1}{\sqrt{2}} \sqrt{1\pm\xi/\sqrt{\xi^2+|\Gamma(\delta)|^2}}\) and \(e^{i\theta}=\Gamma(\delta)/|\Gamma(\delta)|\). The corresponding energies as a function of \(\epsilon\) are shown with red (even) and green (odd) solid lines in Figs.~\ref{fig:theo_gate}a (\(\delta{=}0\)) and \ref{fig:theo_gate}b (\(\delta{=}\pi\)), whose crossings indicate a parity switch of the ground state (vertical black dotted lines). 

When the ancillary state is introduced (\(\ket{\mathcal{E}}_A=\ket{0}_A\), \(\ket{\mathcal{O}}_A = \ket{\sigma}_A\)), for vanishing $t_A$ the many-body energy levels of the whole system \(\ket{M,A}\) are

\begin{align}
	 &\ket{\mathcal{E},\mathcal{E}} \rightarrow E_{\mathcal{E}} & \qquad  & 
      \ket{\mathcal{E},\mathcal{O}} \rightarrow E_{\mathcal{E}} + \epsilon_A \nonumber \\
	 &\ket{\mathcal{O},\mathcal{E}} \rightarrow E_{\mathcal{O}} & \qquad &  \ket{\mathcal{O},\mathcal{O}} \rightarrow E_{\mathcal{O}} + \epsilon_A, 
\end{align}

\noindent
which coincide with those in the main channel if the ancillary level is empty, and are shifted by \(\epsilon_A\) if it is occupied (we consider that the gate voltage in the analyzed range only tunes the main chain level \(\epsilon\) and barely affects \(\epsilon_A\)). These states with the occupied ancillary level switch the global parity with respect to the parity in the main channel, and are indicated with dashed lines in Figs. \ref{fig:theo_gate}a,b. It should be noted that the vanishing coupling of the ancillary level to the superconducting leads renders the resonator quite insensitive to its occupation, thus, the single-tone measurements mainly probe the population in the main channel.

The corresponding global parity conserving transitions of the whole system are shown in Figs. \ref{fig:theo_gate}c,d. As in the two-tone measurements in Fig.~\ref{fig4}, they are shifted (or reflected over \(E{=}0\)) by a constant, which in the model is \(2\epsilon_A\). In addition, they intersect close to the singlet/doublet boundary of the main channel and they may exhibit sharp cusps at its sides, depending on the number of crossings between states with the same global parity  (vertical red/green dotted lines). In order to observe these transitions a finite population in the lowest energy states of both global-parity sectors is needed, and this requires a finite poisoning in the main channel or in the ancillary level depending on the position of the gate.

\begin{figure}[t]
\begin{center}
\includegraphics[width=0.475\textwidth]{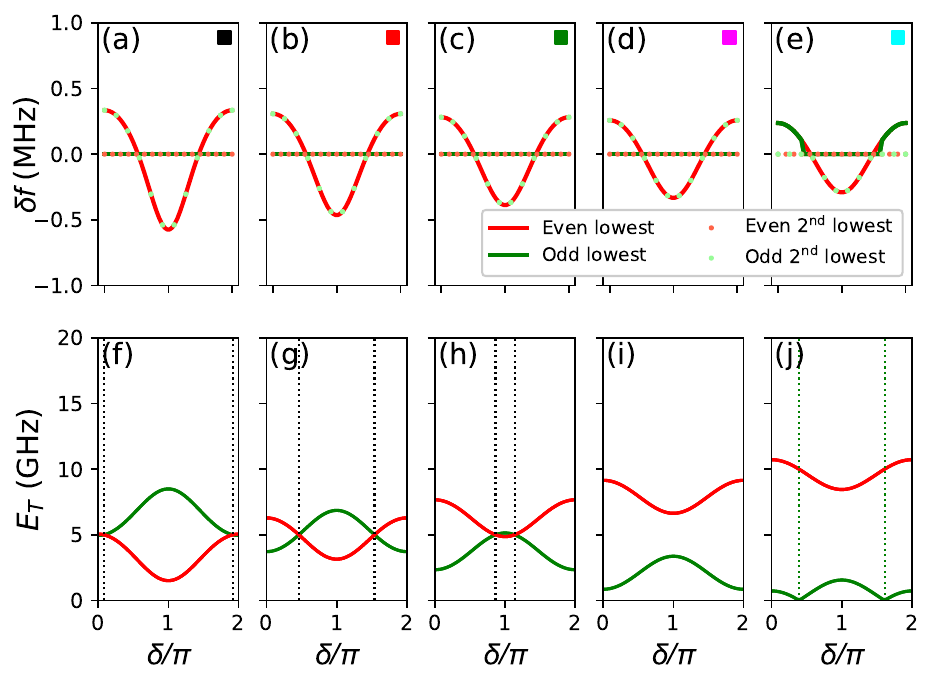}
 \caption{(a-e) Frequency shift over the phase difference for a set of values of the main channel level \(\epsilon\)'s, corresponding to the coloured markers in Fig.~\ref{fig:theo_gate}, following qualitatively those from Figs.~\ref{fig4}, \ref{fig5}. (f-j) Associated lowest global parity conserving transitions. Parameters and vertical lines as in Fig.~\ref{fig:theo_gate}, and the coupling with resonator is set to $\lambda = 0.015$.}
 \label{fig:theo_phase}
 \end{center}
\end{figure}

The phase dependence of the transitions for several values of the main channel level, denoted with color markers in Fig.~\ref{fig:theo_gate}, is shown in Figs.~\ref{fig:theo_phase}(f--j). These results demonstrate that the model can account for the evolution of the experimental transition lines in Fig.~\ref{fig5}(b). First, in Fig.~\ref{fig:theo_phase}(f), $\epsilon$ is placed at the singlet/doublet boundary at \(\delta{=}0\) (see black marker in Fig.~\ref{fig:theo_gate}d) and to the right of the dip in the global even sector at \(\delta{=}\pi\) (black marker in Fig.~\ref{fig:theo_gate}(c)). Next, the level position is raised up until the singlet/doublet boundary at \(\delta{=}\pi\) is almost reached in Fig.~\ref{fig:theo_phase}(h) (green marker in Fig.~\ref{fig:theo_gate}(c)). Finally, the dip in the global odd sector at \(\delta{=}0\) occurs for $\epsilon$ values between those in panels (i) and (j), the latter being placed a bit to the left of the odd dip at \(\delta{=}\pi\). 

The corresponding resonator shift \(\delta f\) for the two lowest levels in each parity sector is shown in Figs.~\ref{fig:theo_phase}(a--e). As discussed above, in the \(t_A\rightarrow0\) limit the shift induced by each state only depends on the main channel, so it disperses with the phase when \(\ket{M}=\ket{\mathcal{E}}\) and is completely suppressed when \(\ket{M}=\ket{\mathcal{O}}\). In order to account for the slight phase dependence of the shift in the odd-like states it is necessary to go beyond this $\Delta \rightarrow \infty$ model, as discussed in the SM \cite{SM}.

In the single-tone spectroscopy, the signal manifests the shifts induced by the states that are significantly populated over the measuring time. Though in general it is expected that most of the population dwells in the ground state, the actual steady state of the junction is determined by processes involving the quasiparticles above the gap and the coupling with the environment \cite{Olivares2014,Ackermann2023}, which induce a non-thermal distribution.\\

\section{Conclusion}
We explored the single-tone and two-tone microwave spectroscopy in a superconducting InAs weak link  close to pinch-off. Observation of jumps in the resonance frequency from the single-tone spectroscopy is understood as singlet--doublet phase transitions that occur due to the reduction of the coupling of the weak link to the superconducting leads. We observed anomalous microwave driven transitions in two-tone spectroscopy, which mimic parity flipping behavior. These parity flip mimicking transitions were tentatively understood as appearing due to the presence of an ancillary level weakly coupled to the weak link.
This behavior might not be generic: we observed it in a single device and it might depend on the particularities of a single device (geometry, defects). However, this shows how the measurement of the excitation spectrum brings crucial information about the system that is not accessible in the ground state properties.
This could be relevant for applications of hybrid structures that require a precise quantum dot configuration, such as Andreev spin qubits implemented in quantum-dot Josephson junctions \cite{PitaVidal2023}. Finally, the parity flipping transitions provide a mechanism to dynamically influence the parity population through the drive, different to the one that involves the continuum of quasiparticles above the gap \cite{Wesdorp2021,Ackermann2023,Olivares2014}.

\section*{Acknowledgments}

Technical support from P. S\'enat is gratefully acknowledged. We thank P. Orfila and S. Delprat for nano-fabrication support, L. Tosi for assistance in data analysis and discussions, C. Metzger, our colleagues from the Quantronics group and G. O. Steffensen for useful discussions. We thank Will Oliver and Lincoln Laboratories for providing us with the TWPA used in the experiment. The nanowire materials development were supported by the European Research Council (ERC) under the grant agreement No. 716655 (HEMs-DAM). M.R.S. benefited from an Investissements d'Avenir grant from Labex PALM (ANR-10-LABX-0039-PALM). A.L.Y. and F.J.M. acknowledges support from the Spanish AEI through grants PID2020-117671GB-I00 and through the “Mar\'ia de Maeztu” Programme for Units of Excellence in R\&D (Grant No. MDM-2014-0377) and from Spanish Ministry of Universities (FPU20/01871). Support by EU through grant no. 828948 (AndQC) is also acknowledged.

\color{black}

\bibliography{references}{}
\end{document}

\title{\texorpdfstring{Supplementary Information \\Ground state phase diagram and ``parity flipping'' microwave transitions \\in a gate-tunable Josephson Junction}{Supplementary Information}}

\maketitle
\nopagebreak[0]

{
\hypersetup{linkcolor=black}
\tableofcontents
}

\newpage

\section{Device and measurement setup}

\begin{figure*}[h!]
\begin{center}
\includegraphics[width=1\textwidth]{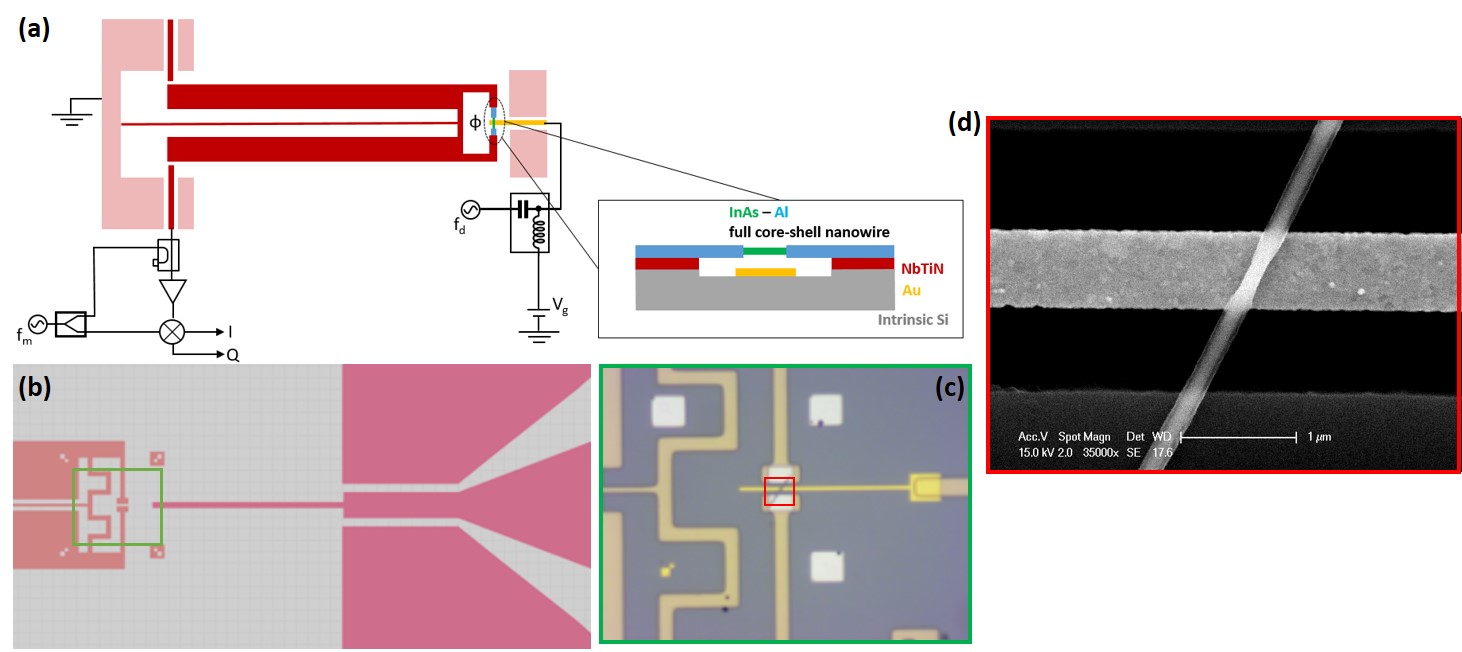}
 \caption{(\textbf{a}) Schematic of the device which consists of a coplanar stripline $\lambda/4$ resonator coupled to an InAs nanowire weak link coupled to a phase-biased superconducting loop through a shared inductance. Measurement tone (frequency $f_m$) through the probe line and drive tone (frequency $f_d$) through the gate line are used to perform the single-tone and two-tone spectroscopy. (\textbf{b}) Design of the device around the phase-biased weak link region. (\textbf{c}) Optical image of the device around the phase biased weak link region. (\textbf{d}) SEM image of the InAs-Al full core-shell nanowire weak link.
}
 \label{figs1}
 \end{center}
\end{figure*}  

Device fabrication starts with the sputtering of a 80-nm-thick NbTiN film on an intrinsic Si wafer at 600$\degree$C.  Then the microwave resonator along with the loop with pads for depositing the nanowire and external parts of the gate electrode are patterned by optical lithography, followed by plasma etching. The etching removes the NbTiN film and creates a 100-nm-deep groove in the substrate. The gate electrode is defined in this groove (e-beam lithography, gold evaporation and lift-off) before transferring the InAs nanowire, so that the nanowire is suspended above the gate. The nanowire is connected to the NbTiN film with thermally deposited  Aluminum patches after Argon etching. In a last step the Aluminum shell around the nanowire is chemically (using Transene D) etched on a short section above the gate electrode, thus forming the weak link. A scanning electron microscope (SEM) image of the weak link is shown in the Fig.~\ref{figs1}(d).\\

\begin{figure*}[b!]
\begin{center}
\includegraphics[width=1\textwidth]{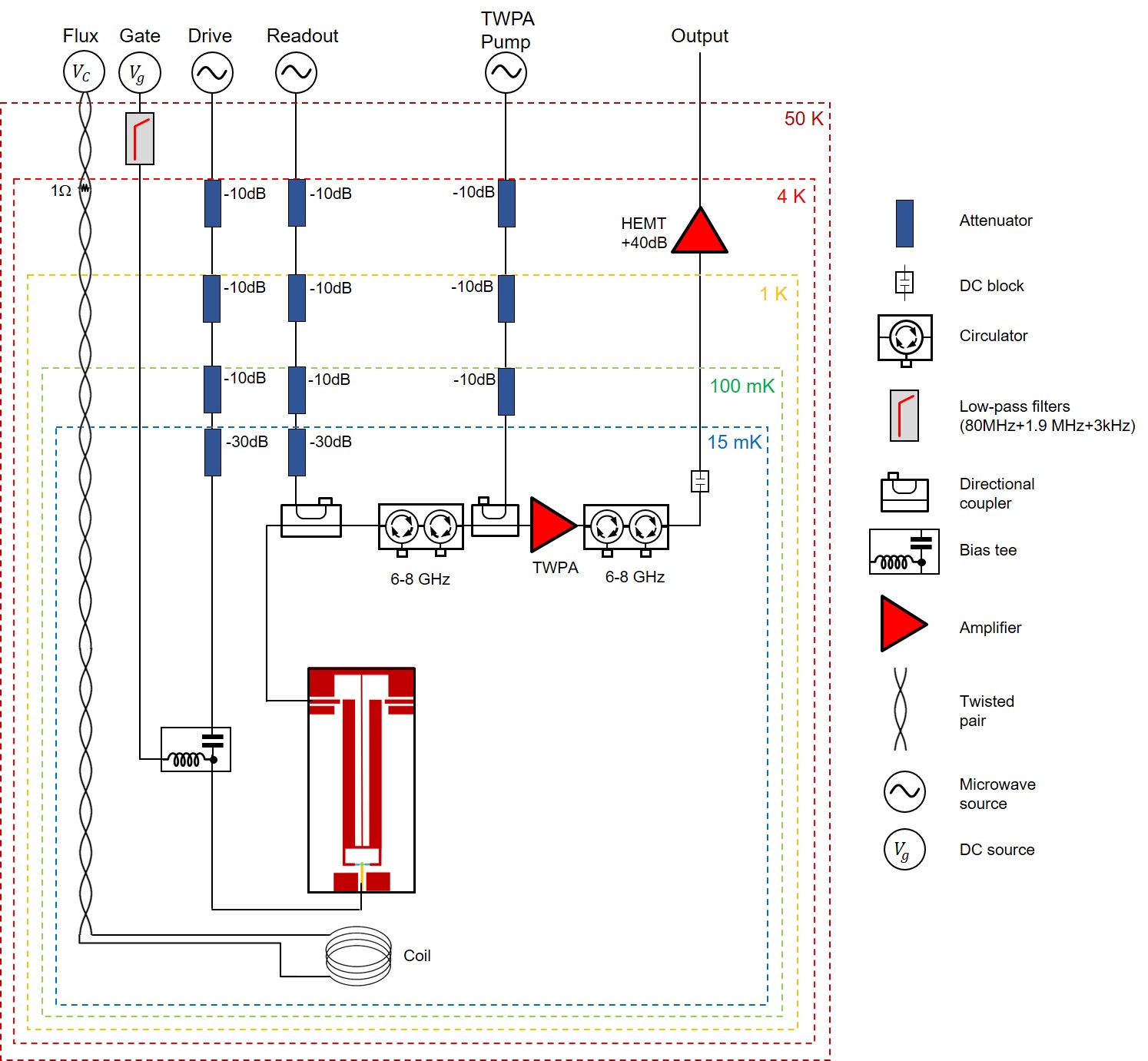}
 \caption{Fridge wiring.
}
 \label{figs2}
 \end{center}
\end{figure*} 

The device is measured in a dilution refrigerator at $\sim$ 15 mK. The measurement (readout) and the drive lines are highly attenuated by series of attenuators to reduce the noise and to efficiently thermalize the central conductors of the microwave cables. The reflected signal is amplified by a combination of travelling wave parametric amplifier (TWPA) with $\sim$ 25dB gain sitting at the MC plate and a low noise amplifier based on high electron mobility transistor (HEMT) with $\sim$ 40dB gain sitting on the 4K plate. A superconducting coil placed around the sample holder allows us to apply a small magnetic field to phase bias the weak link.

\clearpage

\section{Single tone spectroscopy}

\begin{figure*}[h!]
\begin{center}
\includegraphics[width=0.9\textwidth]{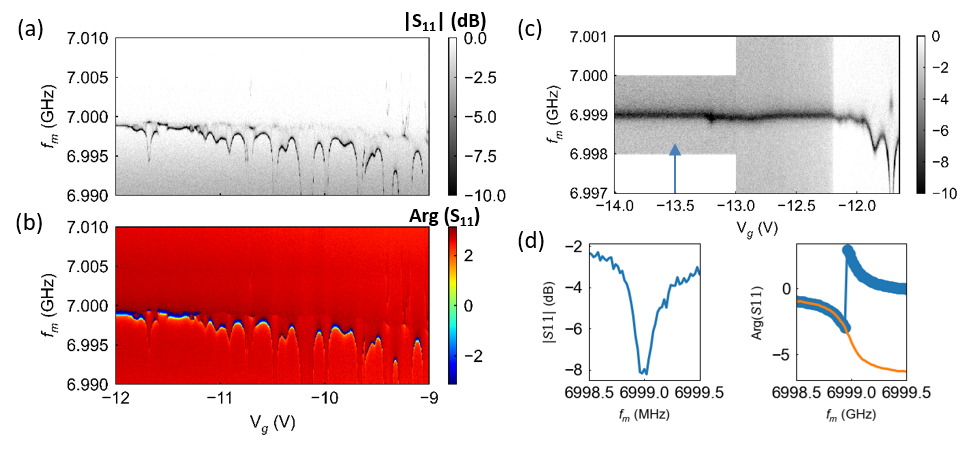}
 \caption{(\textbf{a,b}) 2D color maps of the amplitude and phase of the reflection coefficient $S_{11}$ plotted as a function of gate voltage ($V_g$) and measurement frequency ($f_{m}$). (\textbf{c}) At further negative $V_g,$ pinch-off is seen as the constant response in $V_g$. In this regime, the resonance frequency is $f_{0}$. Note that there is a jump in TWPA gain of $\sim$ 2dB around $V_g=-$ 12.2~V, which is the reason for the sharp change in contrast. (\textbf{d}) Amplitude and phase of $S_{11}$ averaged over 11 scans around the region marked with vertical arrow in (c), plotted as a function of measurement frequency ($f_{m}$) to locate the precise bare resonance frequency: $f_{0}=6.999~$GHz. The orange curve in right panel correspond to the unwrapped phase.
}
 \label{figs3}
 \end{center}
\end{figure*}  

In Fig.~\ref{figs3}(a-b), we plot the continuous wave single tone spectra as a function of gate showing modulations of resonance frequency. To extract the bare resonance frequency $f_{0}$, which is the resonance frequency when the nanowire is completely open circuit, the InAs wire is pinched-off by applying a large negative gate voltage. In Fig.~\ref{figs3}(c) it can be seen that the resonance frequency remains completely gate-independent for voltage below $-13.3$~V, which indicates that the nanowire is then completely pinched-off. In Fig.~\ref{figs3}(d) we plot the amplitude and the phase of $S_{11}$ as a function of $f_{m}$, by taking average over 11 scans around $V_g=-13.5$~V, which shows the resonance $f_0=6.999$~GHz. There is slight variation of TWPA gain, which is the reason for changing background as a function of $f_m$ in Fig.~\ref{figs3}(a). The data of Fig.~2(a) only differ from those of Fig.~\ref{figs3}(a), by the subtraction of a background corresponding to the frequency-dependent TWPA gain.\\

\begin{figure*}[th!]
\begin{center}
\includegraphics[width=1\textwidth]{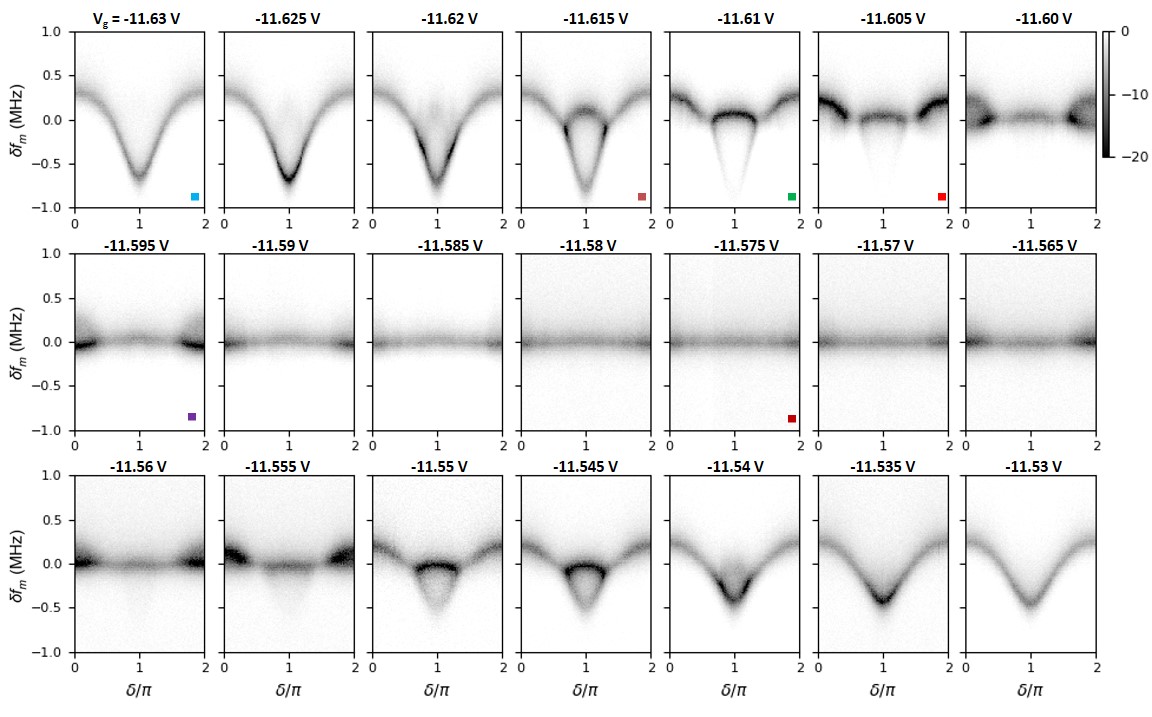}
 \caption{Single-tone spectra as a function of phase difference ($\delta$) at gate voltages between $V_g$ = -11.63 V and  $V_g$ = -11.53 V, in 5 mV steps.
}
 \label{figs4}
 \end{center}
\end{figure*}

\begin{figure*}[h!]
\begin{center}
\includegraphics[width=1\textwidth]{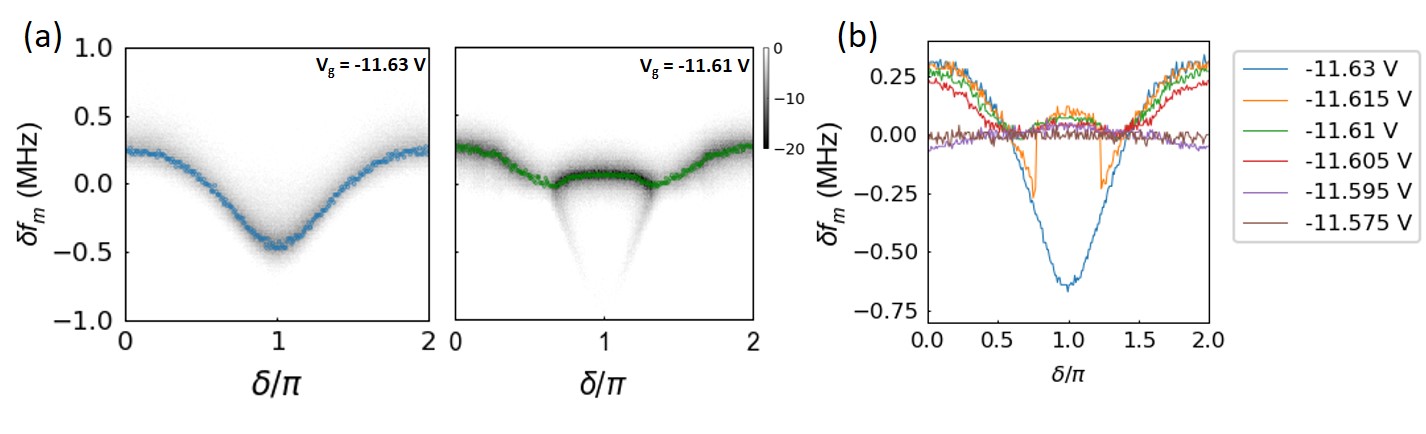}
 \caption{ (\textbf{a}) The frequency of the strongest resonance (or, the global minima of $|S_{11}|(\delta f_m)$) is highlighted on top of the 2D color map of single-tone versus flux. (\textbf{b}) frequency of the strongest resonance plotted as a function of phase difference, for a series of values of $V_g$.
}
 \label{figs5}
 \end{center}
\end{figure*}

In Fig.~\ref{figs4} we plot the single tone spectra as a function of phase difference $\delta$, in the gate range from $-11.63$~V to $-11.53$~V, within which the transitions between the singlet and the doublet ground state happen. At $V_g = -11.53$~V, the resonance frequency disperses in a manner expected for an even ground state with a transition frequency approaching the resonator frequency from above, at $\delta=\pi$. As we keep increasing the gate voltage, first the phase transition to doublet happens at $\delta = \pi$, and the width of the doublet region grows and becomes completely doublet at $V_g = -11.59$~V. More interesting details about the dispersion of the odd ground state can be seen by tracking how the phase dependence of corresponding resonance frequency behaves at different gate voltages. In Fig.~\ref{figs5}(b) we plot such resonance frequency shifts, which manifest stronger dispersion of odd ground state resonance frequency away from center of the doublet region compared to flatter resonance frequency close to $f_0$ in the central region. This, probably, is a situation of minimum supercurrent at the center of the Coulomb blockade.\\

It should be noted that the region over which the singlet - doublet phase transitions are observed was hysteretic in $V_g$ when the  gate sweep is over a large range and there are slight differences between the highlighted region of Fig.~1(b) and the upper panel of Fig.~1(c) in the main manuscript. But, when the gate voltage is swept only in a range of few hundreds of millivolts, the response becomes non-hysteretic and remains stable for weeks.

\clearpage

\section{Evaluation of population from histograms in IQ plane}

\begin{figure*}[b!]
\begin{center}
\includegraphics[width=0.7\textwidth]{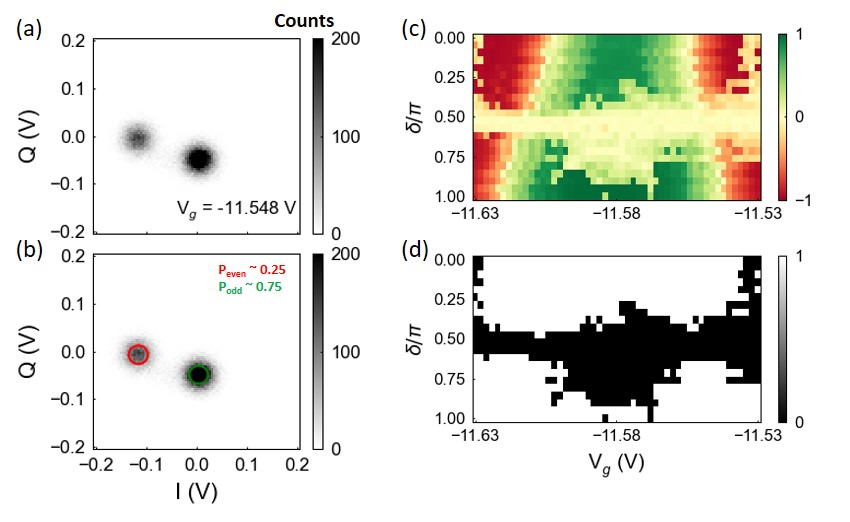}
 \caption{ (\textbf{a}) Histogram of 50000 times measurement of $S_{11}$ at $V_g$ = -11.548 V ($\delta = \pi$) showing the two clouds corresponding to even and odd ground states. (\textbf{b}) GMM fit of the histogram in (a), assuming spherical covariances. The circles correspond to the mean positions and standard variation ($\sigma$). (\textbf{c}) 2D color map of polarization $P_o - P_e$ as a function of gate voltage and phase difference. (\textbf{d}) 2D color map showing whether GMM worked or not: white (ones) meaning GMM works and black (zeroes) meaning GMM fails. Whenever the distance between mean positions of cloud is less than 2$\sigma$, we say GMM fails.
}
 \label{figs6}
 \end{center}
\end{figure*}

The evaluation of the statistical populations of the states of the system results from repetitive measurements of the quadratures $I$ and $Q.$ Each measurement consists of a 218-ns-long high-amplitude (four times the amplitude of the measurement pulse) pre-pulse to rapidly load the resonator with photons, followed by a 500-ns-long measurement pulse. We then plot the histogram of the outcome of 50000 measurements. Such an histogram is plotted in Fig.~\ref{figs6}(a), showing two clouds which correspond to the even and the odd ground states. We use a Gaussian mixture model (GMM) from sklearn python package (sci-kit learn) \cite{scikit-learn}, which employs an expectation-maximization algorithm to find out the best fit, and extract the populations. The GMM fitting of the data in Fig.~\ref{figs6}(a) is shown in Fig.~\ref{figs6}(b), where the centers of the green and red circles are the positions of the clouds centers and their radius the standard deviation $\sigma \sim$ 0.019 V. We then fixed this $\sigma$, and extracted the population at all gate voltages and phase differences, leaving to the GMM the determination of the clouds positions. In Fig.~\ref{figs6}(c), we plot the polarization $P_o - P_e$ as a function of phase difference and gate voltage. GMM fails to identify the clouds in two situations (1) obvious situation when the clouds merge, which is the case around $\delta \sim \pi/2$; (2) sometimes when one of the states is not visible in the histogram. The points where GMM fails are highlighted in blue (zeroes) in Fig.~\ref{figs6}(d). Whenever black points are surrounded with by white points such that $|P_o - P_e| > 0.93$, we set $|P_o - P_e| = 1$ for those black points, yielding Fig.~3(c).\\

\begin{figure*}[h!]
\begin{center}
\includegraphics[width=1\textwidth]{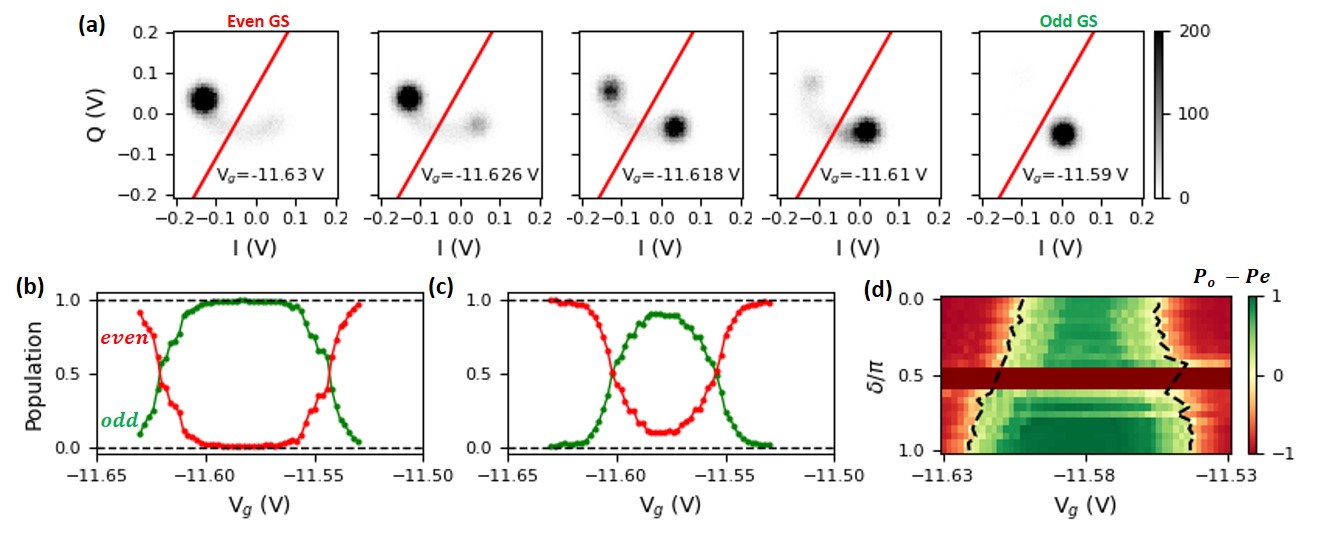}
 \caption{ (\textbf{a}) Histogram of $S_{11}$ measurements in the $IQ$ plane with red line used for states discrimination. (\textbf{b-c}) Population of even and odd states plotted as a function of gate voltage at $\delta = \pi$ and $\delta = 0$, respectively. (\textbf{d}) Corresponding 2D color map of polarization $P_o - P_e$ as a function of gate voltage and phase difference. The black dashed line correspond to zero polarization and the brown region corresponds to situations in which the clouds overlap.
}
 \label{figs7}
 \end{center}
\end{figure*}

We have compared the GMM results with a simpler method consisting in drawing a straight line between the clouds, and count the events on both sides. As the clouds move slightly with changing gate voltage (for a given phase difference), we adjusted progressively the position of the line. In Fig.~\ref{figs7}, we illustrate the method with data taken at $\delta = \pi$, leading to the populations in  Fig.~\ref{figs7}(b). The result at $\delta=0$ are shown in Fig.~\ref{figs7}(c), and the complete phase dependence in Fig.~\ref{figs7}(d). In the region defined by a brown rectangle in Fig.~\ref{figs7}(d), the method could not be applied because the  clouds have a strong overlap. Black dashed lines in Fig.~\ref{figs7}(d) marks the phase boundary ($P_o - P_e=0$) between the doublet and the singlet ground state.

\clearpage

\clearpage

\section{Parity life time from continuous measurement}

\begin{figure*}[h!]
\begin{center}
\includegraphics[width=0.9\textwidth]{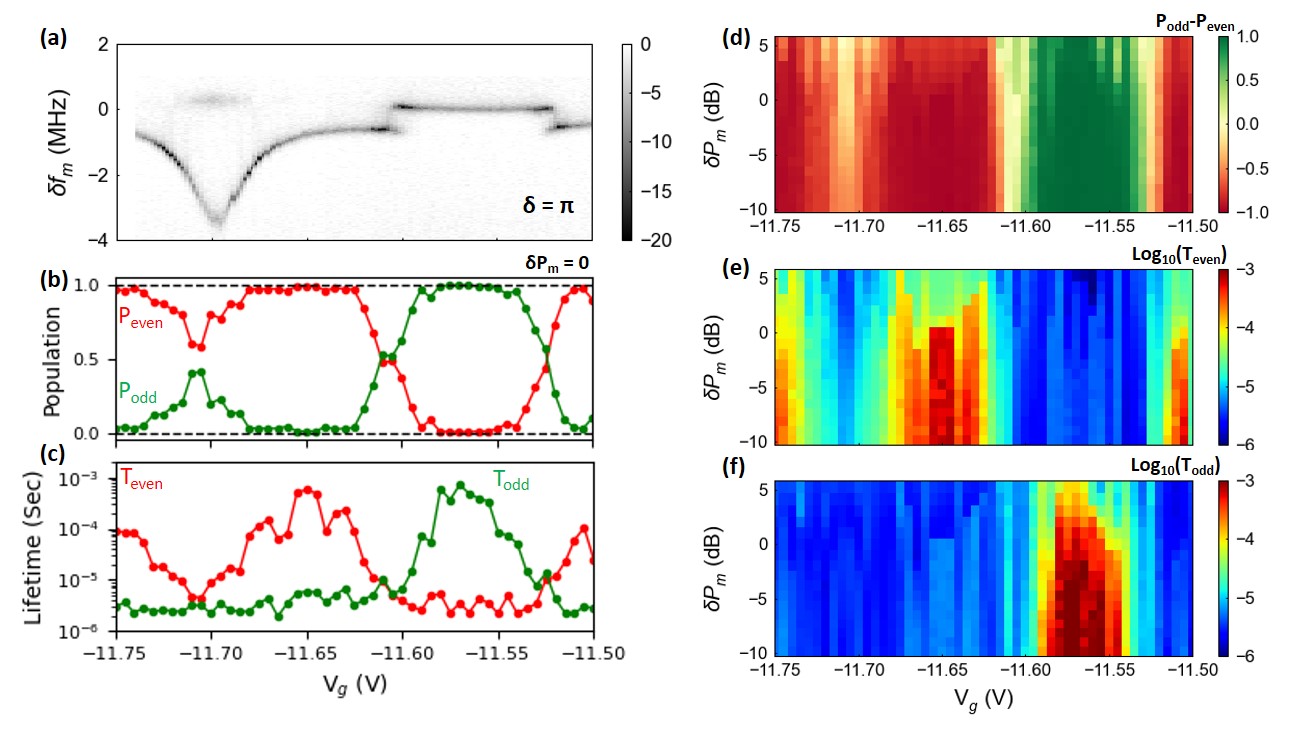}
 \caption{Parity life times and their power dependence at $\delta = \pi$. (\textbf{a}) 2D color map of single-tone spectra as a function of $V_g$. Note that compared to to Fig. 1.(c) and Fig. 3(a), this is another set of data taken after several excursions in $V_g$, which explains a little shift of the position of the jumps in $V_g$. (\textbf{b-c}) Population and parity, respectively, at $\delta P_{Meas}=0$, extracted from continuous measurement. (\textbf{d}) 2D color map of polarization as a function of gate voltage and measurement power.  (\textbf{e-f}) 2D color maps of logarithm of parity lifetimes of even and odd states.
}
 \label{figs8}
 \end{center}
\end{figure*}

The evaluation of parity life time of the two states is inferred from continuous monitoring of the state. The mean value of the quadratures $I$ and $Q$ was recorded on $5\times 10^5$ successive 1~$\mu$s-long intervals. The resulting records was analyzed by a hidden Markov model (HMM) algorithm using the ``SMART'' Matlab package \cite{greenfeld2012single,gammelmark2014hidden}. In Fig.~\ref{figs8}(b) and Fig.~\ref{figs8}(c) we plot the populations and the parity life times, respectively, of the even and odd states as a function of gate voltage measured with a power $\sim$ -124 dBm at $\delta = \pi$. This is the power used for all other histograms presented in the manuscript and SM. Fig.~\ref{figs8}(a) shows the corresponding single-tone spectra. The power dependence of the polarization is shown in Fig.~\ref{figs8}(d), defining the power change $\delta P_m$ relatively to the $-124~$dBm mentioned before. For $\delta P_{m} < 0$ the polarization is independent of power when the population of one of the states is close to 1. The parity time of this state is $\sim 1~$ms, when the other state has a parity life time of few micro seconds or less (the time slots of $1~\mu$s hides faster events). Whenever the polarization is close to zero both the parity life times become few tens of micro seconds and are strongly power-dependent. Similar strong power dependency close to singlet-doublet transition was reported in Bargebos \textit{et al.} \cite{bargerbos2022singlet} and was attributed to parity pumping effect \cite{Wesdorp2021,Ackermann2023}.

\section{Two-tone spectroscopy}

For two-tone spectroscopy, the drive tone is supplied through the gate line via a bias-Tee placed on the mixing chamber plate of the dilution refrigerator. All the features with non-zero $\delta$I/$\delta$Q correspond to some microwave transitions which modifies $S_{11}$. In the main manuscript we focus on two anomalous transitions named as TA and TB, which mimick parity flips. We first show that these are not replicas of some primary transition (T1/T2). 

\clearpage

\subsection{Are the anomalous transitions TA and TB replicas?}

\begin{figure*}[h!]
\begin{center}
\includegraphics[width=0.9\textwidth]{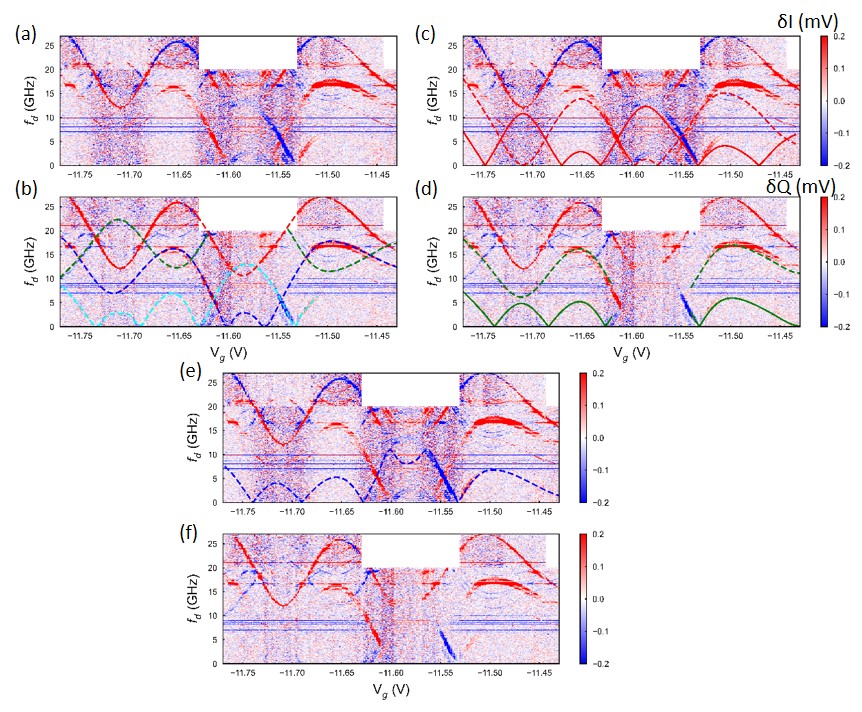}
 \caption{2D color maps of the two-tone spectra as a function gate voltage shown on a broader frequency range. The same data are repeated in (a,c,e) ($\delta I$) and (b,d,f) ($\delta Q$). (\textbf{b}) T1, T2, TA, and TB are highlighted by red, green, blue, and magenta dashed lines, respectively. (\textbf{c}) We tried to compare TA and TB, with T1 shifted in frequency, by equations $TA= |T1-12~$GHz$|$ and  $TB= |T1-23~$GHz$|$, respectively. (\textbf{d}) We tried to compare TA and TB, with T2 shifted in frequency, by equations $TA= |T2-17.5$~GHz$|$ and  $TB= |T2-28.5~$GHz$|$, respectively. Note that, the T2 is not present in the data in the range $-11.62$~V to $-11.54$~V. TA and TB do not have a constant frequency difference with T1 or T2. (\textbf{e})  We tried to compare TB from TA, by equation $TB= |TA-11~$GHz$|$, which somewhat matches.
}
 \label{figs9}
 \end{center}
\end{figure*}

When the drive power is strong enough, one can observe replica of the transition  lines corresponding to processes in which the drive tone excites at the same time an Andreev transition and another fixed mode of the circuit or of the environment, or combine the drive tone and photons from the resonator to excite a transition. From a primary transition at frequency $f$, these processes give rise to spectroscopy lines at $f\pm f_{\rm env},$ with $f_{\rm env}$ the frequency of an environmental mode. In Fig.~\ref{figs9}(c) and (d) we tried to compare the positions of TA and TB, with those of T1 and T2, shifted in frequency. One observes that one cannot reach an overlap on the complete interval. Interestingly, the positions of TB and TA are related by the relation TB= |TA-11~GHz|. However, TA and TB cannot be replica one of the other, because their contrasts are not correlated. The relation between TA and TB finds an explanation in the ancillary level model described below if the level energy does not change with gate voltage. The observation that TA and TB cross very close to the gate voltages where the transitions between singlet to doublet ground tone appear in single-tone can not be a mere coincidence, provided the ancillary level model explaining such alignment.

\clearpage

\subsection{Population transfer by driving TA and TB}

By driving at frequencies corresponding to the transition lines TA and TB, we observe  population transfer between the even and odd cloud, as shown in the histograms in Fig.~\ref{figs10}(d). The general properties we observe about the parity flip mimicking transitions are the following. (1) They generally have a well like shape with minima appearing around the center of odd ground state region and whenever the well crosses zero frequency the part below zero frequency inverts. (2) The inverted part corresponds to odd to even transition, whereas the part that has well like appearance corresponds to even to odd transition. (3) Their crossings points appear close to zero polarization. All the above properties could be explained by an ancillary level model.

\begin{figure*}[h!]
\begin{center}
\includegraphics[width=1\textwidth]{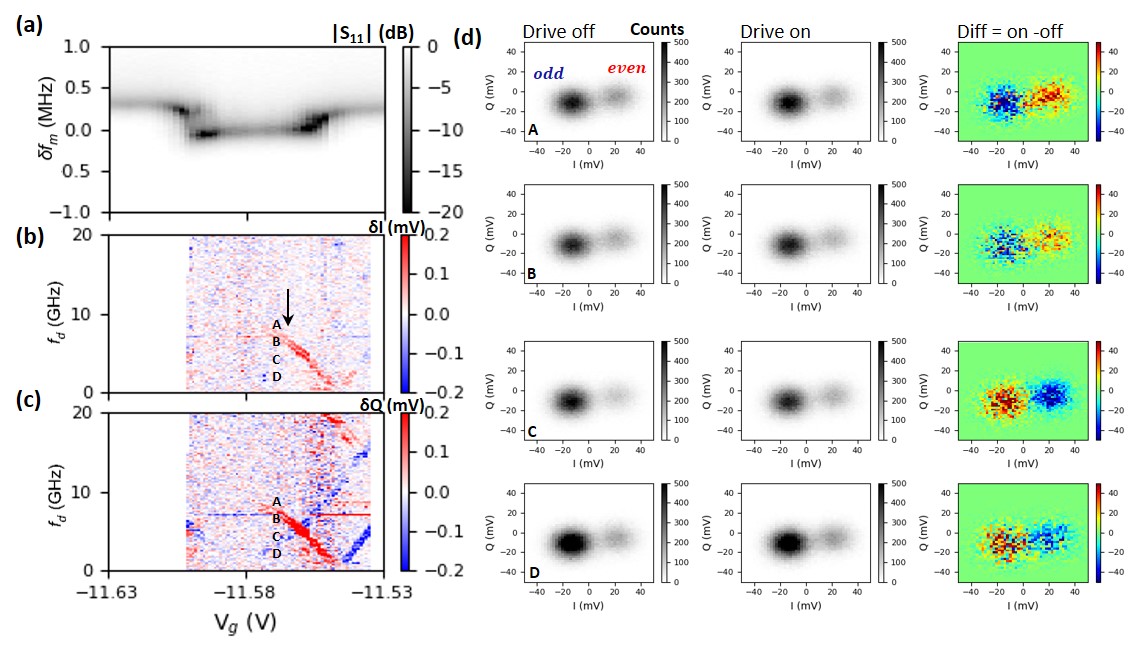}
 \caption{ (\textbf{a}) Other set of single-tone data around the singlet-doublet transition in gate voltage measured with a slightly higher power compared to previously shown measurements. (\textbf{b-c}) Corresponding two-tone data showing the anomalous transitions TA and TB. (\textbf{d}) Histograms of $S_{11}$ measurements in IQ plane when the drive is off and on are plotted at the drive frequencies marked by A-D in (b) and (c). The difference of histogram shows that the well shaped part of TA and TB corresponds to even to odd transition, and, the inverted well shaped part corresponds to odd to even transition.
}
 \label{figs10}
 \end{center}
\end{figure*} 

\clearpage

\subsection{Rotation of IQ plane to put most of the signal in one component}

The IQ plane is rotated by a flux dependent angle $\Theta$ such that when $f_d$ lies on TA $\delta I \hat{x}+\delta Q \hat{y}$ aligns with the new $a$-axis, which help putting most of the signal in one component.

\begin{figure*}[h!]
\begin{center}
\includegraphics[width=0.6\textwidth]{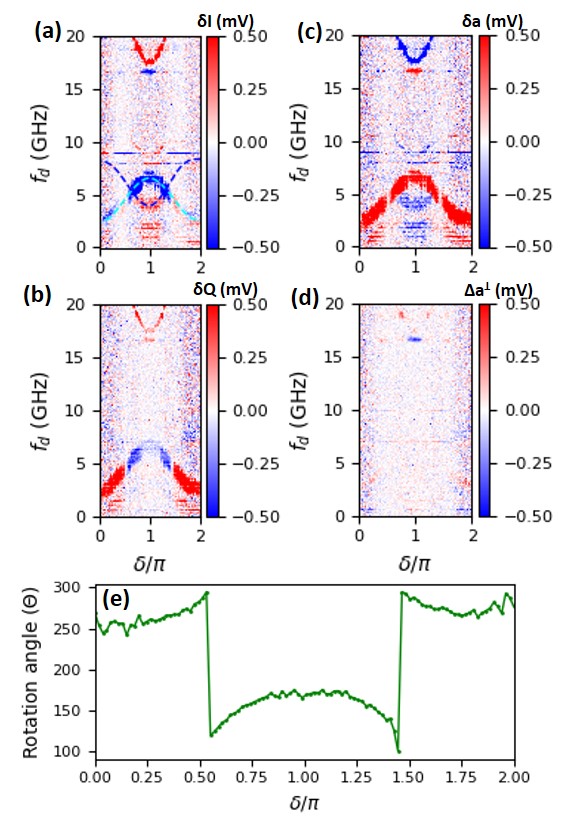}
 \caption{ (\textbf{a-b}) Two tone data $\delta I$ and $\delta Q$ at $V_g =$ -11.55 V. (\textbf{c-d}) The rotated two tone data $\delta a$ and $\delta a^{\perp}$, such that most of the signal lies in $\delta a$. (\textbf{e}) The rotation angle plotted as a function of flux. The jump of $\pi$ around $\delta = \pi /2$ correspond to the flux when the resonance frequency corresponding to even-like state crosses the resonance frequency corresponding to odd-like state.
}
 \label{figs10}
 \end{center}
\end{figure*} 

\clearpage

\section{Theory}

\subsection{Zero bandwidth model}

The weak link in the regime analyzed in this work is modeled with a main channel \(M\) and a an ancillary level \(A\). In the main text, \(M\) is described with the infinite gap limit of a single level Anderson model, where the effect of the superconducting leads is integrated into an effective pairing \(\Gamma(\delta)\) between the even states of a dot \cite{Meng2009}. While this approach already accounts for the qualitative structure of the experimental measurements, a direct improvement of the description of the superconducting leads is to use the zero-bandwidth (ZBW) approximation \cite{martin2011josephson,Affleck2000}, by which each lead is modeled with one superconducting site:

\begin{align*}
    H_{WL} &= H_M + H_A \\
    H_M &= \sum_{\sigma} \epsilon d^\dagger_\sigma d_\sigma + U n_\uparrow n_\downarrow 
    + \left( \sum_{i=L,R} \Delta_i c^\dagger_{i\uparrow} c^\dagger_{i\downarrow} + h.c. \right) 
    + \left( \sum_\sigma t_{L}e^{-i\delta/4} c^\dagger_{L\sigma} d_\sigma + t_{R}e^{i\delta/4} d^\dagger_\sigma c_{R\sigma} + h.c. \right) \\
    H_A &=  \sum_{\sigma} \epsilon_A d^\dagger_{A\sigma} d_{A\sigma}  
	+ U_A n_{A\uparrow} n_{A\downarrow} 
    + \sum_\sigma U_{MA} n_{\sigma} n_{A\bar{\sigma}} 
    + \left( \sum_\sigma  t_A d^\dagger_{A\sigma} d_{\sigma} + t_{LA} e^{-i\delta/4} c^\dagger_{L\sigma} d_{A\sigma} + t_{RA}e^{i\delta/4} d^\dagger_{A\sigma} c_{R\sigma} + h.c. \right),
\end{align*}

\noindent
where \(t_{L,R}\) are the tunnel amplitudes between the main dot and the leads, \(\Delta_{L,R}\) are the gaps of the leads, \(t_A\), (\(t_{LA},t_{RA}\)) are the vanishing tunnel amplitudes between the ancillary level and the main level (the leads), and \(U_{MA}\) is an interlevel charging energy that penalizes the simultaneous occupation of \(M\) and \(A\). 

 \begin{figure}[b]
\includegraphics[width=1\textwidth]{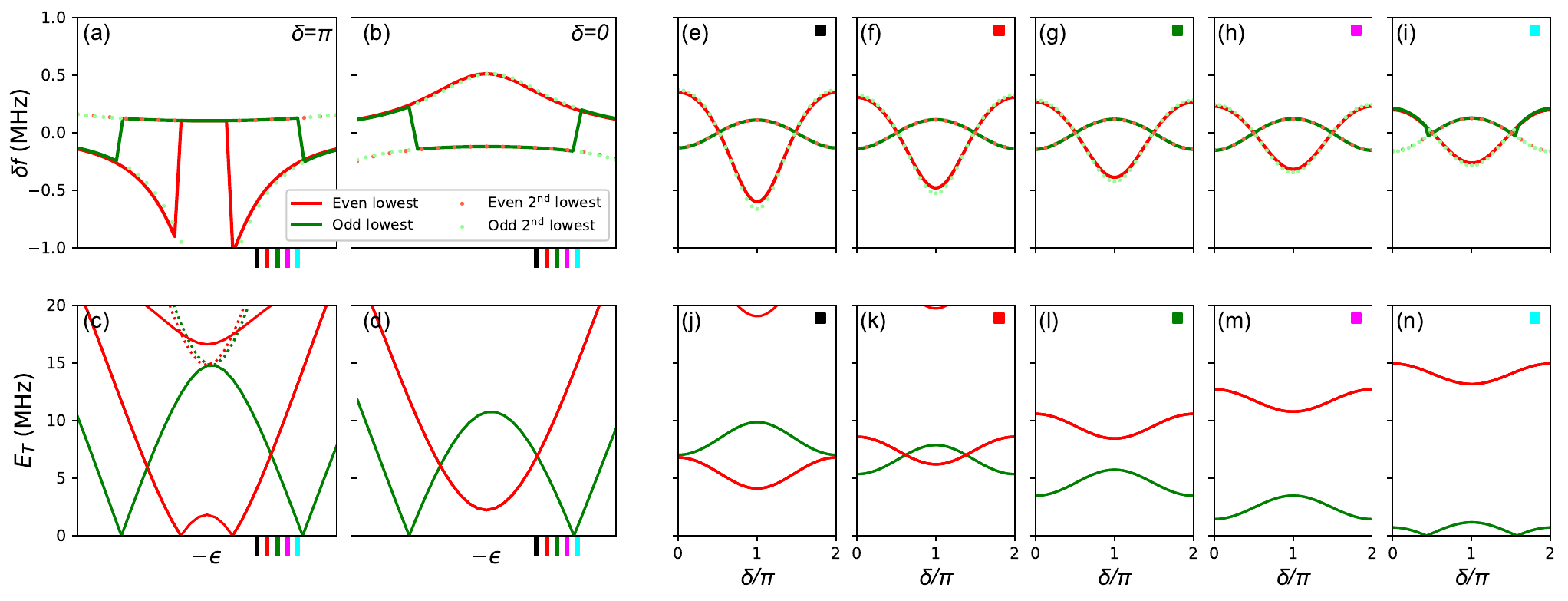}
 \caption{ Zero bandwidth model. (a,b) Frequency shift over \(\epsilon\) at \(\delta=0,\pi\). (c,d) The corresponding transitions of global even (red) and odd (green) parity. Solid (dashed) lines switch (maintain) the ancilla state. (e-i) Frequency shift over \(\delta\) at \(\epsilon\)'s marked with colored labels in (c,d). (j-n) The corresponding transitions. Parameters are: $t_L=5$, $t_R=10$, $\Delta_{L,R}=40$, $\epsilon_A=5$, $U=55$, $U_A=1.5$ (units in GHz), and $\lambda=0.02$.
}
 \label{fig:SM_ZBW}
\end{figure}

The resonance frequency of the resonator coupled to the weak link depends on the state \(n\) of the latter. At moderate coupling \(\lambda\), the resulting shift \(\delta f_n\) from the bare resonance frequency \(f_0\) can be calculated with \textcolor{red}{\cite{Park2020,Metzger2021}}:

\begin{equation}
    \delta f_n = f_n-f_{0} = \lambda^2 \bra{\Phi_n} \frac{\partial^2 H_{WL}}{\partial \delta^2} \ket{\Phi_n} - \lambda^2 \sum_{m\neq n} |\bra{\Phi_m} \frac{\partial H_{WL}}{\partial \delta} \ket{\Phi_n}|^2 \left( \frac{1}{E_m-E_n + hf_0} + \frac{1}{E_m-E_n - hf_0} \right) 
     ,
\end{equation}

\noindent
where \(H_{WL}\) describes the weak link, and \(E_n\), \(\ket{\Phi_n}\) are the corresponding energies and states. The second term, which corresponds to the exchange of virtual excitations between the junction and the resonator (dispersive shift), provides a small contribution to the shift in our case. This occurs because the current matrix elements involved in the transitions TA and TB that cross the resonator line (\(E_m{-}E_n = hf_0\)) are negligible. Thus, the main contribution to the dispersive shift is produced by the pair transition, which is a couple of GHz higher than the resonator. As shown in Figs.~\ref{fig:SM_ZBW}a,b,e-i, the ZBW model allows to recover a finite phase dispersion in the odd state of \(M\). 

The corresponding global parity conserving transitions between Andreev states are shown in Figs.~\ref{fig:SM_ZBW}c,d,j-n. As in the main text, the solid lines with cusps flip the parity in the ancillary level and in the main channel, switching between the lowest energy state in each parity of $M$ (in the infinite gap approximation they are $\ket{g}_M= u\ket{0}_M - v e^{i\theta}\ket{\uparrow\downarrow}_M$, that was denoted with $\ket{\mathcal{E}}_M$ in the main text, and $\ket{\sigma}_M$, that was denoted with $\ket{\mathcal{O}}_M$). Here, we additionally plot higher transitions -- which are also present in the infinite gap model but were not discussed nor shown. The second solid red line flips the parity in $A$ and switches between the lowest odd and the second lowest even state in $M$ (in the infinite gap approximation the latter one is $\ket{e}_M = v\ket{0}_M + u e^{i\theta}\ket{\uparrow\downarrow}_M$). Analogously to the previous set, this line has a companion in the global odd sector at a $2\epsilon_A$ upward shift, which is above the range of the graph. The dotted lines are transitions that do not flip the parity in $A$ and go from $\ket{g}_M$ to $\ket{e}_M$, namely, the {\it pair} transitions in the main channel produced at different occupation of $A$, which may be associated with T1. Note that a finite \(U_{MA}\) is included, which breaks the mirror symmetry of the ensemble of lines over the point \(\epsilon = -U/2\) (\(\xi=0\)).

\subsection{Multilevel dot model}

An alternative development of the weak link model is to consider additional levels in the main channel in order to describe a larger gate range. Proceeding within the infinite gap limit,
the main channel Hamiltonian updates to:

\begin{align*}
    H_M = &\sum_{\alpha\sigma} \epsilon_\alpha d^\dagger_{\alpha\sigma} d_{\alpha\sigma} + \sum_{\alpha} U_\alpha n_{\alpha\uparrow} n_{\alpha\downarrow} + 
    \sum_{\alpha\beta\sigma} t_{\alpha\beta} d^\dagger_{\alpha\sigma} d_{\beta\sigma} + \\
    & \sum_{\alpha} \left( \left(\Gamma_{L\alpha} e^{i\delta/2} + \Gamma_{R\alpha} e^{-i\delta/2} \right) d^\dagger_{\alpha\uparrow} d^\dagger_{\alpha\downarrow} + h.c. \right) + 
    \sum_{\alpha \neq \beta} \left( \left(\Gamma_{L\alpha\beta} e^{i\delta/2} + \Gamma_{R\alpha\beta} e^{-i\delta/2} \right) d^\dagger_{\alpha\uparrow} d^\dagger_{\beta\downarrow} + h.c. \right),
\end{align*}

\noindent
where \(\epsilon_\alpha\), \(U_\alpha\) are the position and the charging energy of the dot levels $\alpha$'s, \(t_{\alpha\beta}\) is a tunnel amplitude between two of them, \(\Gamma_{\alpha L(R)}\) is an effective pairing from each lead $L$($R$) onto each level, and \(\Gamma_{L(R) \alpha\beta }\) is an effective interlevel pairing that describes processes where a Cooper pair from the lead splits into levels \(\alpha\) and \(\beta\), which is chosen as $\sqrt{\Gamma_\alpha\Gamma_\beta}$. As in the previous models, tunneling terms between the main levels and the ancillary one are supposed to be small, and interlevel charging terms of any kind are not included for simplicity.  

\begin{figure}[t]
\includegraphics[width=1\textwidth]{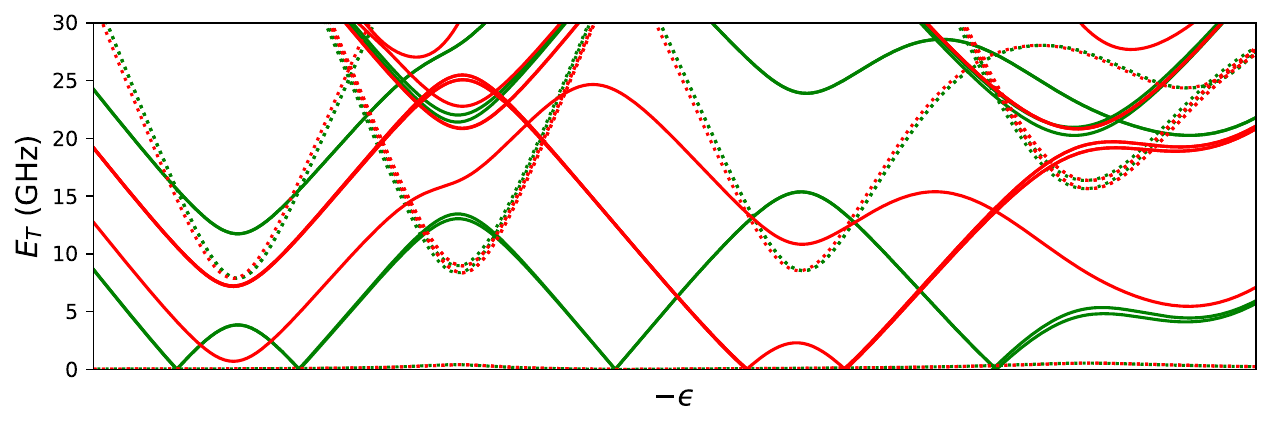}
 \caption{Transition lines within the multilevel dot model (sketched in the inset) at $\delta=\pi$ and sweeping $\epsilon$. Red (green) denotes even (odd) global parity, and solid (dashed) represents lines that switch (maintain) the ancilla state. Parameters are: $\epsilon_A{=}5$, $\epsilon_1{=}50$, $\epsilon_1{=}50$, $\epsilon_2{=}-15$, $\epsilon_3{=}-60$, $\Gamma_{1L}{=}2$, $\Gamma_{1R}{=}15$, $\Gamma_{2L}{=}6$, $\Gamma_{2R}{=}10$, $\Gamma_{3L}{=}2$, $\Gamma_{3R}{=}6$, $U_1{=}1$, $U_2{=}30$, $U_3{=}5$, $U_{1A}{=}U_{2A}{=}U_{3A}{=}0.5$, $t_{12}{=}1$, $t_{23}{=}10$  (units in GHz).}
 \label{fig:SM_ML}
\end{figure}

Transitions calculated with this model are shown in Fig.~\ref{fig:SM_ML}, where three levels are included. The gate dependence is modeled with the same lever arm in each level except in the ancillary one (\(\epsilon_\alpha = \epsilon^0_\alpha + \epsilon\)), and interlevel tunneling and pairing are only present between adjacent levels. The transitions displayed are those starting from the lowest states in each global parity (red/green for even/odd) and each occupation in $A$ (solid/dashed for $\ket{0}$/$\ket{\sigma}$). The {\it anomalous} transitions TA and TB are reproduced as in the previous models in the range where they both exhibit cusps. The other two levels of the dot also display this kind of transitions, centered at the points where $\epsilon_\alpha \approx -U_\alpha/2$, which correspond to the gate points $V_g\sim-11.2$V and foreseeably $V_g\sim-11.42$V in {\it e.g.} Fig. \ref{figs9}. There is only one more pair of cusps displayed by TB in the first gate point, a feature that can be associated to a disfavored odd state (lower $U_\alpha$ or higher $\Gamma_\alpha$) in these dot levels. The connection between the TAs and TBs of the levels is mediated by the neighbouring tunnel couplings $t_\alpha$ and the interlevel effective pairings $\Gamma_{\alpha\beta}$. In these intermediate gate regions they display splittings as in the measurements, though other splittings in the regions where $\epsilon_\alpha = -U_\alpha$ are not reproduced. The other set of solid lines with minima around $\epsilon_\alpha = -U_\alpha$ can be associated, close to these points, to the transitions that switch between $\ket{\sigma}_{M\alpha}$ and $\ket{e}_{M\alpha}$, as discussed in the previous subsection about the zero bandwidth model. Analogously, dashed lines with with minima around $\epsilon_\alpha = -U_\alpha/2$ can be associated, close to these points, to the {\it pair} transitions at different occupations of $A$, so they may be associated with T1. The other set of dashed lines that disperse opposite to the {\it pair} line, with minima between the points where $\epsilon_\alpha = -U_\alpha/2$, can be associated with transitions where a Cooper pair is split into two of the dot levels (in the limit without pairing, it corresponds to a transition involving $\ket{\sigma,\sigma'}$ and $\ket{0,0}$, which intersect at $\epsilon=(\epsilon^0_\alpha+\epsilon^0_\beta)/2$, being $\alpha$ and $\beta$ two consecutive dot levels and $\ket{\alpha,\beta}$ their states). These might be related with T2. We do not attempt to cover other lines in the measurements within our simplified models, which might arise from the intricate combination of geometry, spin-orbit, electron-electron interactions, {\it etc.} in this kind of devices \cite{Tosi2019,Metzger2021,matute2022signatures,Wesdorp2023}.

The parameters for the figure are based in those used for the single-level infinite gap model in the main text (caption of Fig. 6), which are chosen to reproduce certain points in the transition lines: at \(\xi=0\) and \(\delta=\pi\), the pair transition T1 of \(\sim11\)~GHz corresponds to \(2\Gamma(\pi) = 2|\Gamma_L-\Gamma_R|\), and the transitions TA, TB, of \(\sim 2.5, 12.5\)~GHz, to \(-\Gamma(\pi) + U/2 \mp \epsilon_A\), so \(U, \epsilon_A\) are determined. Then TA or TB at \(\delta=0\), of \(\sim 2.5, 7.5\)~GHz, that correspond to \(\mp(\Gamma(0) + U/2 \mp \epsilon_A)\), allows to obtain \(\Gamma(0)=\Gamma_L+\Gamma_R\), which together with \(\Gamma(\pi)\) determines \(\Gamma_{L,R}\). 

\bibliography{references_SM}{}